\documentclass[]{spie}  
\usepackage{amsmath,amsfonts,amssymb}
\usepackage{graphicx}
\usepackage{setspace}
\usepackage{tocloft}
\usepackage{float}
\usepackage{siunitx}
\usepackage{subcaption}
\usepackage[table]{xcolor}
\usepackage{url,hyperref}
\DeclareSIUnit\bit{bit}
\DeclareSIUnit\byte{B}
\DeclareSIUnit\rad{rad}
\usepackage{multirow}
\usepackage{multicol}
\usepackage{tabularx}
\newcolumntype{Y}{>{\centering\arraybackslash}X}
\usepackage[nohypertypes={acronym}]{glossaries}
\glsdisablehyper

\newacronym{gcr}{GCR}{galactic cosmic ray}
\newacronym{spe}{SPE}{solar particle event}
\newacronym{tid}{TID}{total ionizing dose}
\newacronym{see}{SEE}{single-event effect}
\newacronym{seu}{SEU}{single-event upset}
\newacronym{sel}{SEL}{single-event latchup}
\newacronym{sefi}{SEFI}{single-event functional interrupt}
\newacronym{soi}{SOI}{silicon-on-insulator}
\newacronym{gaa}{GAA}{gate-all-around}
\newacronym{cots}{COTS}{commercial off-the-shelf}
\newacronym{tmr}{TMR}{triple modular redundancy}
\newacronym{dmr}{DMR}{dual modular redundancy}
\newacronym{ecc}{ECC}{error-correcting code}
\newacronym{abft}{ABFT}{algorithm-based fault tolerance}
\newacronym{dm}{DM}{deformable mirror}

\usepackage{pifont}
\definecolor{darkgreen}{RGB}{0,165,0}
\newcommand{\cmark}{\textcolor{darkgreen}{\checkmark}}
\usepackage{xcolor} 
\newcommand{\cross}{\textcolor{red!100!black}{\ding{55}}}
\usepackage{array}
\newcolumntype{P}[1]{>{\centering\arraybackslash}p{#1}}

\newcommand{\greenup}{\textcolor{darkgreen}{$\uparrow$}}
\newcommand{\greendown}{\textcolor{darkgreen}{$\downarrow$}}
\newcommand{\redup}{\textcolor{red!100!black}{$\uparrow$}}
\newcommand{\reddown}{\textcolor{red!100!black}{$\downarrow$}}

\title{Compute System Organization for High Frequency High Order Wavefront Sensing and Control}

\author[a]{Barry Lyu}
\author[a]{Vaibhavi Manjarekar}
\author[a]{Nathaniel Bleier}
\affil[a]{University of Michigan, Computer Science and Engineering, 2260 Hayward St, Ann Arbor, MI 48109, United States}

\cftpagenumbersoff{figure}
\cftpagenumbersoff{table} 
\begin{document} 
\maketitle

\begin{abstract}
Maintaining long-term wavefront stability is critical for the Habitable Worlds Observatory (HWO), which targets contrasts approaching 
$10^{-10}$ and therefore requires continuous dark-zone maintenance using high-order wavefront sensing and control (HOWFSC). Prior work has advanced HOWFSC algorithms and profiled candidate implementations on radiation-hardened processors, highlighting a substantial gap between the computational demands of LUVOIR-scale HOWFSC and the capabilities of current onboard spacecraft hardware. In this paper, we argue that this gap can be closed by offloading the HOWFSC pipeline to a dedicated co-flying compute satellite at Sun–Earth L2. This approach enables the use of modern, radiation-tolerant high-performance processors without increasing risk to the primary observatory. We show that such an architecture can increase the end-to-end control cadence from the sub-hertz regime typical of radiation-hardened onboard processing or ground-in-the-loop operation to tens and even hundreds of hertz. We evaluate commercial hardware platforms in terms of performance and feasibility, and we propose custom architectures that enable higher control frequencies with significant power consumption reductions. Finally, we outline system-level considerations for co-flying compute, including reliability, satellite integration, and inter-satellite communication constraints.

\end{abstract}

\keywords{Habitable Worlds Observatory; coronagraph; wavefront control; dark hole maintenance; compute systems}

{\noindent \footnotesize\textbf{*}Address all correspondence to Barry Lyu,  barrylyu@umich.edu }

\begin{spacing}{2}   

\section{Introduction}
The Habitable Worlds Observatory (HWO) is a planned NASA flagship mission intended to launch in the early 2040s with the goal of directly imaging and characterizing potentially habitable exoplanets. Achieving this objective requires detecting planets that are at least $10^{10}$ times dimmer than their host stars.  This is enabled by a high-contrast coronagraph system\cite{EFF_DRIFT}. Maintaining such extreme contrast over long durations demands continuous correction of high-order wavefront errors caused by thermal and mechanical drift, making wavefront stability a mission-critical requirement. 

High-order wavefront sensing and control (HOWFSC) provides a path to maintaining a stable dark zone by repeatedly estimating the electric field and updating deformable-mirror commands, making the achievable control cadence a key system parameter. Prior works have developed and profiled HOWFSC algorithms on embedded and space-relevant hardware, emphasizing both the dominant computational kernels and the uncertainty in achievable performance at mission scale\cite{COMPLEX,EMBEDDED,THESIS}. However, the compute and memory demands of Large Ultraviolet Optical Infrared Surveyor (LUVOIR) scale dark-zone maintenance remain far beyond what is practical for current radiation-hardened (RadHard) spacecraft processors, and ground-in-the-loop operation is fundamentally constrained by communication latency and downlink capacity. Motivated by these limitations, we propose offloading HOWFSC to a dedicated co-flying compute satellite equipped with radiation-tolerant high-performance hardware that is not rad-hard by design, preserving the reliability posture of the primary observatory while enabling substantially higher processing capability close to the telescope. We evaluate both compute and satellite-level design options for such offloading, and show that co-flying compute is a promising path to closing the HOWFSC compute gap for HWO.

\subsection{High-Order Wavefront Sensing and Control Algorithm} 

High-order wavefront sensing and control (HOWFSC) algorithms estimate residual starlight in the image plane and compute corrective commands for deformable mirrors \cite{DZM_OG,DZM_IMPL}. Prior work has introduced multiple techniques across Jacobian computation, electric field estimation, and control synthesis, as summarized in Tab. \ref{tab:algos}. 

\begin{table}[t]
\begin{center}       
\begin{tabular}{l|l} 
\hline
\rowcolor{gray!20}\rule[-1ex]{0pt}{3.5ex} \textbf{Purpose} & \textbf{Algorithm} \\
\hline
\multirow{4}{*}{Controls calculation} & \rule[-1ex]{0pt}{3.5ex} Electric field conjugation (EFC)\cite{EFC_CITE}  \\
   & \rule[-1ex]{0pt}{3.5ex} Linear dark-field control (LDFC)\cite{LDFC}  \\
  & \rule[-1ex]{0pt}{3.5ex} Adjoint EFC\cite{ADJOINT} \\
\hline
\multirow{3}{*}{Electric field estimation} & \rule[-1ex]{0pt}{3.5ex} Pairwise probing (PWP) \cite{PWP} \\
  & \rule[-1ex]{0pt}{3.5ex} Extended Kalman filter (EKF) \cite{DZM_OG}  \\
  & \rule[-1ex]{0pt}{3.5ex} Self-coherent camera (SCC) \\
  & \rule[-1ex]{0pt}{3.5ex} Modal EKF \cite{MODAL}\\
\hline
\multirow{3}{*}{Jacobian Creation} & \rule[-1ex]{0pt}{3.5ex} Optical model propagation \cite{OPTICAL_PROP}  \\
  & \rule[-1ex]{0pt}{3.5ex} Expectation maximization (EM)  \\
\hline
\end{tabular}
\end{center}
\caption{Summary of HOWFSC algorithms adapted from Eickert and Pogorelyuk et al. \cite{COMPLEX,THESIS} } 
\label{tab:algos}
\end{table} 

Dark hole maintenance routines with HOWFSC typically leverage a Jacobian matrix to describe the relationship between \gls{dm} actuators and image plane pixels and exhibit significant computational complexities. 


While such HOWFSC algorithms are usually tested in terms of their end-to-end contrast performance, recent studies have highlighted the significant uncertainty and scale of their computational requirements \cite{COMPLEX, THESIS}.  HOWFSC exhibits significant demands for both computational throughput and memory capacity, scaling with the total number of pixels and actuators. For a LUVOIR-class coronagraph, maintaining contrast may require closing the HOWFSC loop at
sub-second timescales, exceeding the capabilities of radiation-hardened (RadHard) hardware by three to five orders of magnitude.

HOWFSC algorithms exhibit a high degree of parallelism and scale well to multiple compute nodes (Section \ref{subsec:parallel}). Multiple on-board RadHard compute can collaborate to achieve a higher frequency, but overcoming the compute gap requires a tremendous amount of nodes, which leads to prohibitive power generation and heat dissipation requirements. These additional compute on-board the telescope may also further aggravate the thermal instabilities of coronagraph instruments. 

The Nancy Grace Roman Space Telescope (ROMAN), scheduled for 2027, will present the first in-space demonstration of a coronagraph with HOWFSC technologies and targets $10^{-8}$ contrast. It bypasses HOWFSC compute requirements with ground-in-the-loop (GITL) HOWFSC\cite{ROMAN_DEMO}. This approach of offloading compute to Earth requires downlinking images to ground and uplinking controls to the telescope. While offloading to Earth effectively unlocks unbounded compute resources, the latency incurred by the sheer distance between Earth and L2 prevents HOWFSC from reaching a high frequency, the round-trip light time from Earth to L2 is approximately \SI{10}{\second}, and the transmission rates across such a large distance are quite low. 

The HWO, which is projected to have a LUVOIR-sized coronagraph, has considerably more actuators and pixels. It exhibits at least 400 times higher computational requirements than ROMAN for a single HOWFSC iteration. Its higher contrast target also mandates HOWFSC control cadences at a much higher frequency, extending beyond the capabilities of some of the most advanced compute systems.

\begin{figure}[t]
\begin{center}
\begin{tabular}{c}
\includegraphics[width=0.6\linewidth]{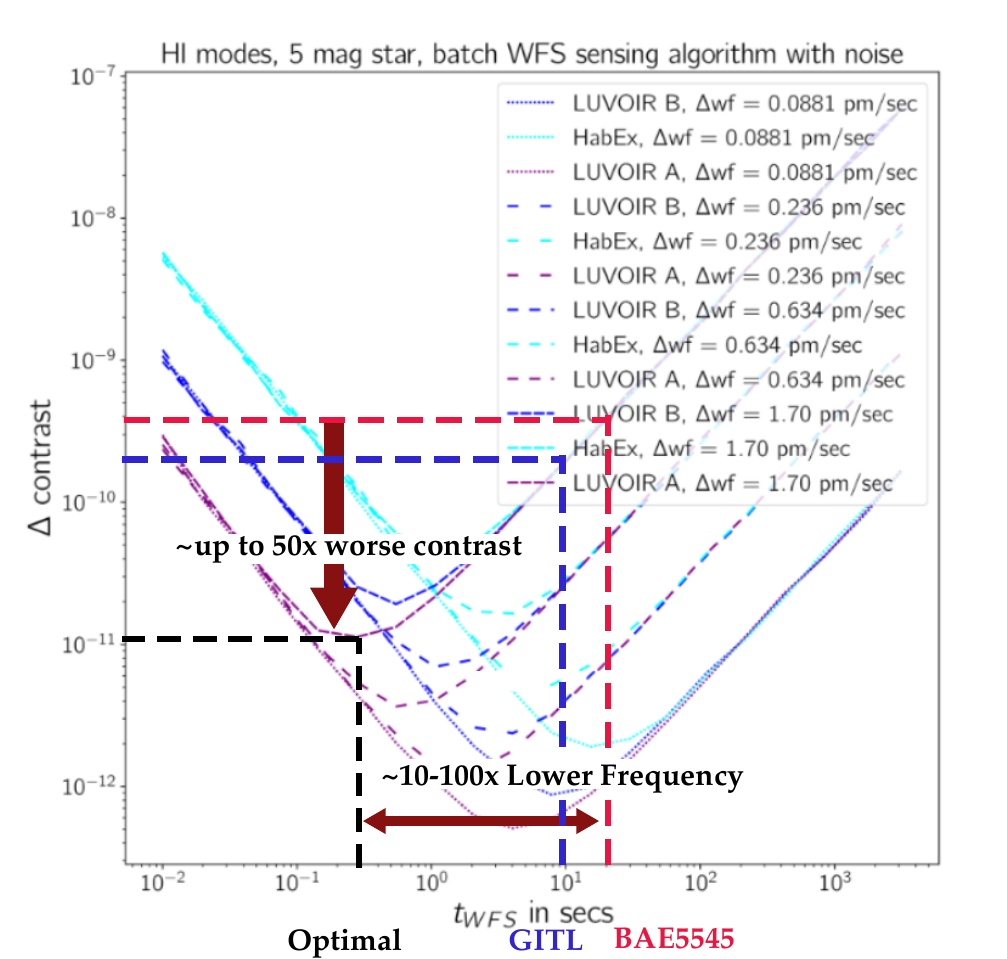}
\end{tabular}
\end{center}
\caption 
{ \label{fig:howfsc_motiv}
Impact of wavefront sensing cadence on contrast by Pueyo et al \cite{PUEYO_PRES}, annotated with achievable cadence and contrast of GITL and on-board BAE5545 for the LUVOIR A system with $\Delta \text{wf}=$\SI{1.70}{\pico\meter\per\second}.}
\end{figure} 

Figure \ref{fig:howfsc_motiv} shows that the optimal control cadence for LUVOIR-A may lie between several tens and hundreds of hertz. In contrast, ground-in-the-loop operations and RadHard onboard processors are limited to cadences that are orders of magnitude lower. Operating at these reduced frequencies can degrade the achievable contrast by up to two orders of magnitude, with correspondingly significant reductions in the exoplanet detection yield and overall mission capability.


\section{Offloading Compute to Co-flying Spacecraft}
The dilemma between low-frequency GITL and prohibitive on-board costs of HOWFSC calls for alternative solutions. Prior studies have hinted at the possibility of co-flying spacecraft, but have yet to comprehensively analyze this strategy \cite{COMPLEX}.  

Co-flying compute offers two primary advantages. First, the hardware constraints on a smaller companion spacecraft are significantly less restrictive than those on the main observatory, which are limited to RadHard processing. A dedicated compute satellite can accommodate radiation-tolerant hardware with much higher performance, as well as custom architectures tailored for HOWFSC with substantially improved energy efficiency. Second, spacecraft that co-locate the Sun-Earth L2 orbit can be orders of magnitude closer to the telescope than the Earth. At these distances, inter-satellite links can provide low latency and high data rates\cite{CUBESAT}, enabling higher control cadences than are possible with ground-based processing.

A dedicated compute satellite would cost only a small fraction of the flagship observatory and could be designed around radiation-tolerant high-performance hardware without the stringent mass, thermal, and reliability constraints of the main telescope. Continued advancements in in-space high-performance computing may further reduce cost and increase capability.

The broader landscape also supports this trajectory. High-performance space computing is a rapidly growing sector. Nvidia and Starcloud announced concepts for gigawatt-scale orbital data centers \cite{starcloud}. Google’s Project Suncatcher proposed similar large-scale satellite compute constellations \cite{SUNCATCHER}. Starlink’s V2 Mini satellites, which already provide substantial on-orbit compute capability, have demonstrated build costs of only \$800,000 per unit \cite{STARLINK_DISR}. These trends suggest that high-performance in-space compute will become increasingly accessible, reliable, and cost-effective in the coming decade. Therefore, the scientific impact of deploying a low-cost, replaceable co-flying compute node for HWO may outweigh its cost. By enabling the high frequency control required for extreme contrasts, such a system has the potential to significantly increase the number of detectable exoplanets, improve the characterization quality, and ensure that the full scientific promise of HWO can be realized.

\begin{figure}[t]
\begin{center}
\begin{tabular}{c c}
\includegraphics[width=0.5\linewidth]{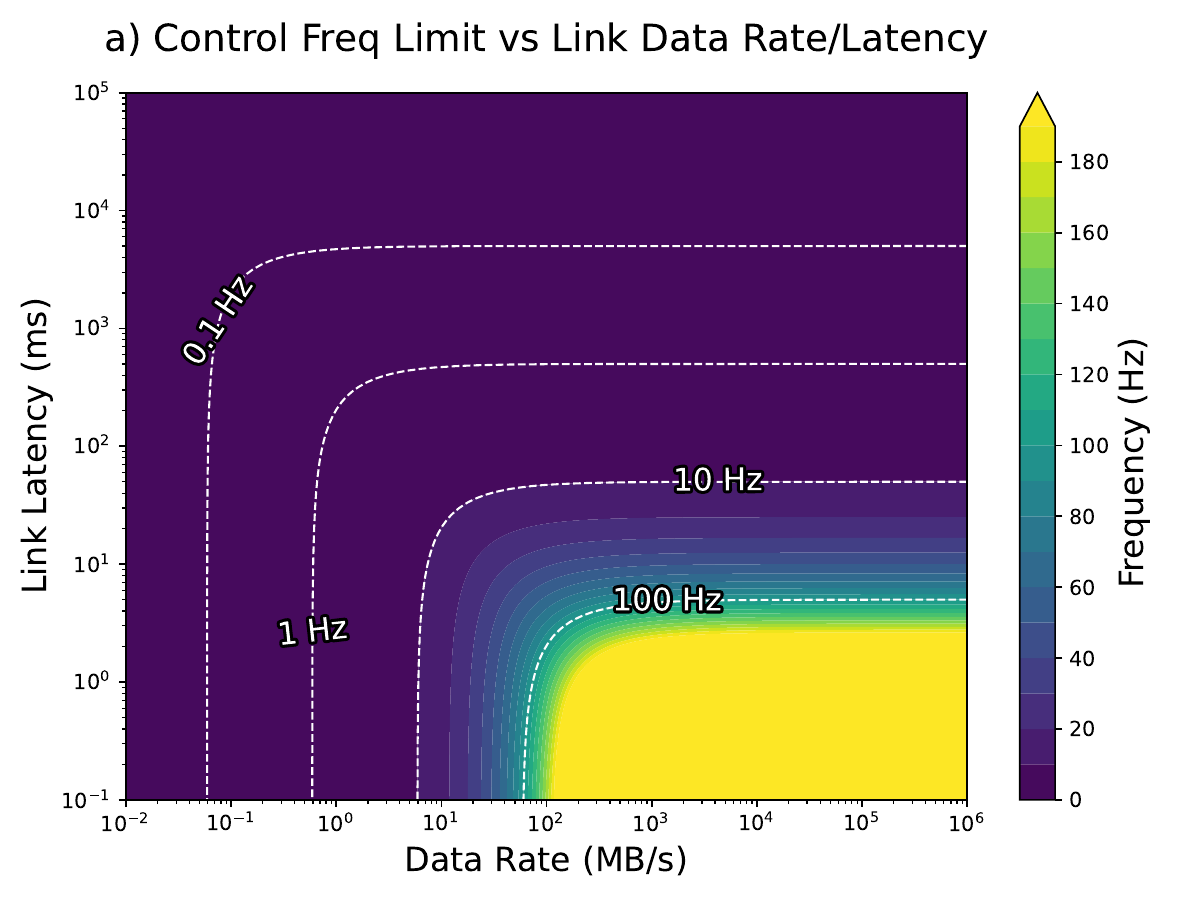} \includegraphics[width=0.5\linewidth]{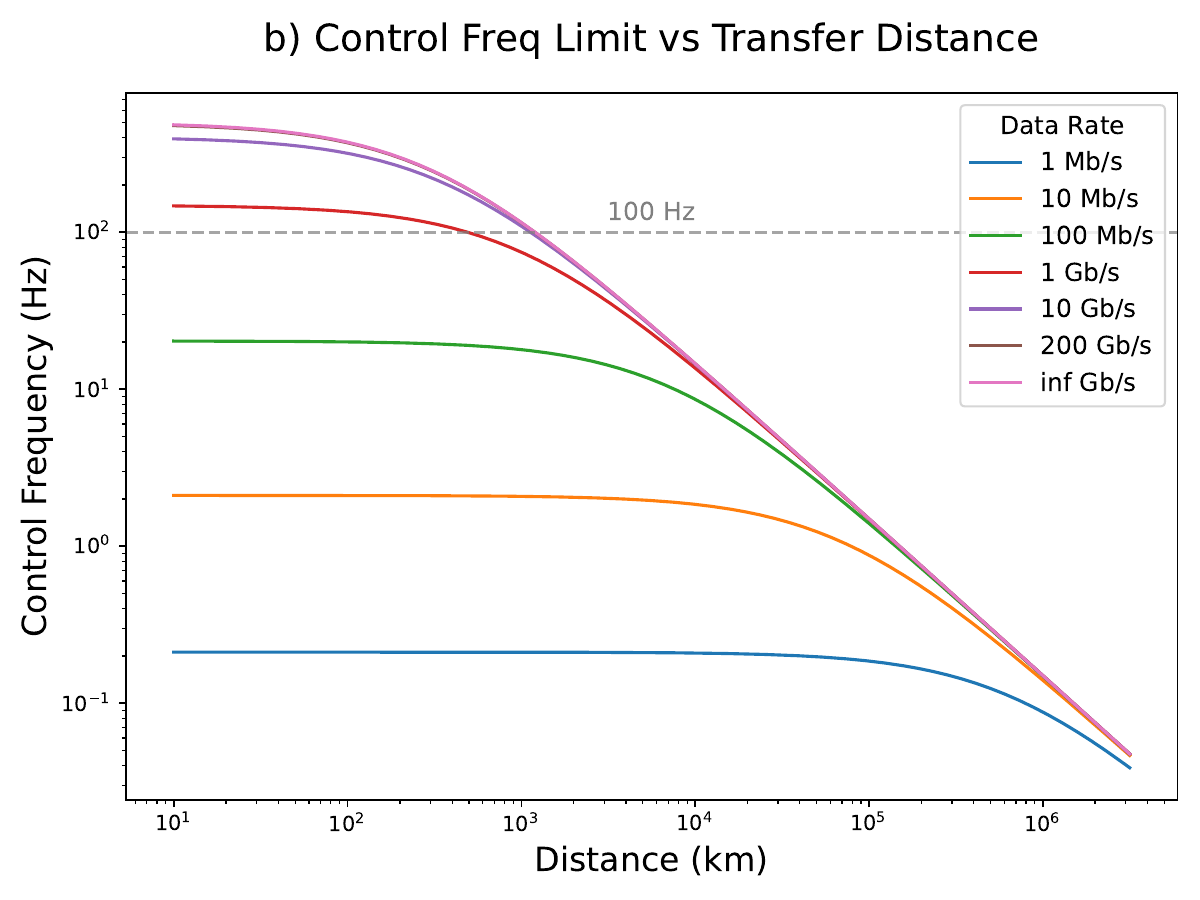}
\end{tabular}
\end{center}
\caption 
{ \label{fig:transfer}
\textbf{a)} Upper-bound of control frequency relative to link latency and bandwidth, assuming infinite compute, symmetric spatial links, and dithering-based estimation (pairwise probing will have higher demand on link bandwidth, scaling with the number of probes). Frequencies are capped to \SI{200}{\hertz} for better visibility. \textbf{b)} Upper bound of control frequency relative to offload distance for various link bandwidths, assuming \SI{1}{\milli\second} latency for signal processing on top of propagation latency. } 
\end{figure} 

Figure~\ref{fig:transfer} show that link latency and bandwidth impose a hard upper limit on the achievable control frequency, regardless of algorithmic improvements or available compute power. In a ground‐in‐the‐loop (GITL) architecture, round‐trip light‐time and downlink constraints restrict control updates to well below \SI{0.1}{\hertz}, far short of the optimal cadence for an HWO‐scale coronagraph.\cite{COMPLEX} As a result, even perfect algorithms and unlimited ground compute cannot compensate for physical communication bottlenecks.

We evaluated the achievable control frequency as a function of offload distance across a range of link data rates, from \SI{1}{\mega\bit\per\second} Deep Space Network Grade~2 service\cite{DSN} to demonstrated \SI{200}{\giga\bit\per\second} free-space optical links\cite{CUBESAT}, as well as a hypothetical infinite-bandwidth case. The results show that high-frequency HOWFSC is only possible when computation is located near the telescope; increasing data rates yield rapidly diminishing returns compared to reducing propagation delay, confirming propagation delay as the dominant constraint for offloaded architectures using common optical and RF communication systems. The analysis further indicates that the control frequency does not degrade meaningfully until separations exceed \SI{300}{\kilo\meter}. This separation range is compatible with proximity operations anticipated for serviceable mission concepts as well as with non-contact co-flying architectures, without imposing tight formation-flying constraints.

In addition to computation and communication latency, the achievable control frequency is bounded by the intrinsic mechanical response of micro-electromechanical system (MEMS) deformable mirrors. Contemporary MEMS DMs typically support update rates in the kilohertz regime, with mechanical settling times ranging from hundreds of microseconds to a few milliseconds depending on the actuator design, stroke, and drive conditions \cite{MEMS_PERF,MEMS_HCI}. While this response time imposes a physical limit on how quickly wavefront corrections can be realized, it exceeds the bandwidth-limited control frequencies considered in this study. Accordingly, we do not explicitly model MEMS dynamics, and instead focus on computation and communication as the dominant constraints on achievable end-to-end update rates in the regimes evaluated.

\section{Analyzing Performance of HOWFSC Algorithms}
Prior studies have thoroughly analyzed and benchmarked HOWFSC algorithms on RadHard processors and field programmable gate arrays (FPGAs), suggesting that orders of magnitude speedup must be attained to lower the HOWFSC latency\cite{COMPLEX, EMBEDDED, THESIS}. We extend these works to analyze the bottlenecks and performances of HOWFSC on modern hardware, leveraging the roofline model, which has been widely adopted in the computer architecture community for performance modeling and software-hardware co-design\cite{ROOFLINE}.

\subsection{Pairwise Probing}
In HOWFSC, pairwise probing serves as a field estimator that converts science-camera photon counts into an estimate of the complex focal-plane electric field. This method applies a small number of known DM probe commands and forms difference images between positive and negative probe pairs, which removes the incoherent intensity term and yields a linear measurement of the coherent electric field. For a given pixel and wavelength, stacking the difference measurements across $N_{\mathrm{PWP}}$ probes gives
\begin{singlespace}
\begin{equation}
\boldsymbol{\delta}
=
2
\begin{bmatrix}
-\Im\{\Delta \mathbf p_1\} & \Re\{\Delta \mathbf p_1\} \\
\vdots & \vdots \\
-\Im\{\Delta \mathbf p_{N_{\mathrm{PWP}}}\} & \Re\{\Delta \mathbf p_{N_{\mathrm{PWP}}}\}
\end{bmatrix}
\begin{bmatrix}
\Re\{E_s\} \\
\Im\{E_s\}
\end{bmatrix},
\end{equation}
\end{singlespace}
where $\Delta \mathbf p_n$ denotes the focal-plane field induced by the $n$-th probe. The real and imaginary components of the electric field were then recovered by solving a small regularized least-squares problem, as shown in Eq. \ref{eq:PWP}, which is performed independently for each pixel and wavelength.
\begin{singlespace}
\begin{equation}
\label{eq:PWP}
\begin{bmatrix}
\Re\{E_s\} \\
\Im\{E_s\}
\end{bmatrix}
=
(\mathbf M^T \mathbf M + \mu \mathbf I)^{-1} \mathbf M^T \boldsymbol{\delta},
\end{equation}
\end{singlespace}

\subsection{Extended Kalman Filter}
Alternatively, EKF with dithering can be employed as the field estimator. It injects phase diversity by adding small random DM perturbations to actuator commands. The single pixel EKF update formulation is given by Eq. \ref{eq:efk_update}, where for a pixel $(i,j)$ at iteration $k$, $\hat E_{ij}(k)$ is the electric field estimation, $K_{ij}$ is the Kalman gain, $\hat{y}_{ij}(k)$ is the predicted photon count, and $\mathbf{J}_{ij}$ is the corresponding vector in the Jacobian\cite{DZM_OG}. 
\begin{singlespace} 
\begin{align}
\label{eq:efk_update}
\begin{bmatrix}
\Re\{\hat E_{ij}(k+1)\} \\
\Im\{\hat E_{ij}(k+1)\} \\
\hat I_{ij}(k+1)
\end{bmatrix}
=
\begin{bmatrix}
\Re\{\hat E_{ij}(k)\} \\
\Im\{\hat E_{ij}(k)\} \\
\hat I_{ij}(k)
\end{bmatrix} +
K_{ij}(k)\,
\big(y_{ij}(k)-\hat{y}_{ij}(k)\big)\\
\hat y_{ij}(k)=\max\!\big(\hat I_{ij}(k), I_D\big)+\left\lvert \hat E_{ij}(k) + \mathbf{J}_{ij}\,\Delta \mathbf{u}(k) \right\rvert^2
\end{align}
\end{singlespace}

We note that while the state updates for each pixel are independent, the Kalman gain $K_{ij}(k)$ and predicted photon count $\hat{y}_{ij}$ both depend on $\mathbf{J}_{ij}\Delta \mathbf{u}(k)$, and the perturbations induced by the control command at the previous iteration $k$. Performing the state update across all pixels thus requires computing the full influence $\mathbf{J}\Delta\mathbf{u}(k)$. We extract this computation as a general matrix-vector multiplication (GEMV), computed once at the beginning of every iteration and stored in memory to avoid redundant computation. Because this influence is only used in the subsequent iteration, for HOWFSC on a co-flying satellite, this computation can also overlap with inter-satellite communication to accelerate the control cadence. 

\subsection{EFC with Precomputed Gain}
EFC takes as input electric field intensity estimations and computes corrective DM commands. For EFC during dark hole maintenance, the time-critical quantity that changes every control iteration is the estimated focal-plane electric field $E$. In contrast, the Jacobian $\mathbf{J}$ and regularization parameter $\alpha$ evolve on much longer timescales, typically set during commissioning and updated intermittently. The expensive linear algebra associated with these algorithms can be performed offboard and then reused across many high-rate iterations. Pogorelyuk et al. \cite{COMPLEX} outlined the formulation for EFC with the precomputed gain matrix $M$, as shown in Eq. \ref{eq:efc_precomp}. All further discussions assume storing the Jacobian, along with the precomputed $M$ matrix in memory. 
\begin{align}
\label{eq:efc_precomp}
\Delta \mathbf u &= \underbrace{-(\mathbf{J}^T\mathbf{J}+\alpha\mathbf I)^{-1}}_M \mathbf J^T\mathbf E = -M\mathbf J^T\mathbf E
\end{align}

\subsection{Modal EKF}
Modal Extended Kalman Filter (Modal EKF) is a recently proposed approach for dark zone maintenance that estimates wavefront drift on a reduced-order modal basis rather than on a per-pixel basis. Instead of treating each focal-plane pixel as an independent state, the Modal EKF represents the open-loop electric field as a linear combination of physically meaningful modes, typically aligned with the DM basis. By exploiting the global structure of the wavefront, this formulation improves the estimation efficiency under photon-limited conditions and approaches theoretical contrast limits more closely than single-pixel EKF methods \cite{MODAL}. Equation~\ref{eq:HK} presents the formulation of the observation matrix $H_k \in \mathbb{R}^{n \times m}$, while Eq.~\ref{eq:KK} describes the computation of the Kalman gain $K_k$.
\begin{align}
\label{eq:HK}
\mathbf{H}_k &= 4\,\mathbf{B}\,\mathrm{diag}\!\left(\mathbf{J}\,\delta\mathbf{u}_p\right)\,\mathbf{J}\\
\label{eq:KK}
\mathbf{K}_k &= \mathbf{P}_{k}\,\mathbf{H}_k^T\left(\mathbf{H}_k\,\mathbf{P}_{k}\,\mathbf{H}_k^T+\mathbf{R}_p\right)^{-1}
\end{align}
The full-state Modal EKF shown in Eqs.~\ref{eq:HK} and \ref{eq:KK} is very computationally expensive, as it involves multiple large matrix multiplications and exhibits $\mathcal{O}(\texttt{\#}\text{Actuator}^3)$ complexity. Applying full-state Modal EKF to LUVOIR-scale dark holes at high control frequencies may require exascale levels of compute power, which far exceeds the energy and power budgets of deep-space spacecraft and is therefore more suitable for ground-in-the-loop operation. To address this limitation, a reduced-rank formulation based on singular value decomposition (SVD) could lower computational cost by retaining only the $r$ most significant modes instead of the full actuator space. This reduced rank approach opens up opportunities for space-based lower latency operation, but the optimal compute architecture depends heavily on the number of ranks. Due to remaining uncertainties in performance, robustness, and implementation overhead, we leave a comprehensive characterization and evaluation of these reduced-rank approaches to future work. 

\subsection{Roofline Model Analysis}
The roofline model characterizes an algorithm based on operational intensity, as defined by Eq. \ref{eq:OI}. It then models the attainable throughput of that algorithm on a particular processor, defined by Eq. \ref{eq:ROOF}, according to the processor's peak floating-point performance and memory bandwidth. Recognizing that the above HOWFSC kernels are predominantly dense linear algebra, the roofline model is an excellent fit for characterizing their performance requirements.
\begin{align}
    \label{eq:OI}
    \text{Operational Intensity} &= \frac{\text{Number of floating point arithmetic operations}}{\text{Number of bytes accessed from main memory}}\\
    \label{eq:ROOF}
    \text{Attainable Throughput} &= \min\begin{cases} 
    \text{Peak Floating-Point Performance} \\
    \text{Peak Memory Bandwidth} \times \text{Operational Intensity} 
    \end{cases}
\end{align}

\begin{figure}[t]
\begin{center}
\begin{tabular}{c}
\includegraphics[width=0.6\linewidth]{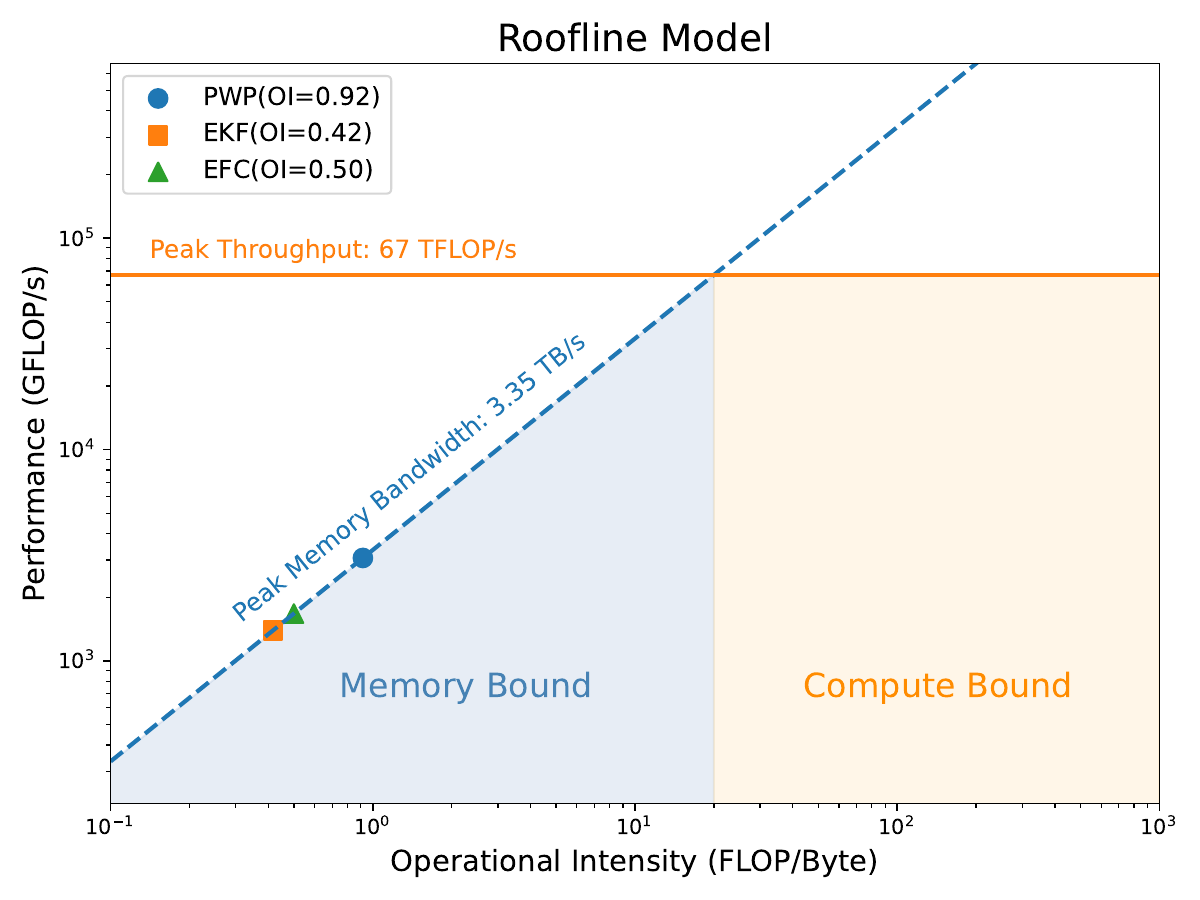}
\end{tabular}
\end{center}
\caption 
{ \label{fig:ROOF}
Roofline plot of three key dark hole maintenance kernels of a LUVOIR system on the Nvidia H100 GPU, assuming precomputed Jacobian and inverse matrices stored in single precision.} 
\end{figure}

Figure \ref{fig:ROOF} maps three key dark hole maintenance kernels of a LUVOIR-scale system to the Nvidia H100 GPU. All three algorithms have a very low operational intensity, which places them in the memory-bound region of the H100 GPU. This indicates that the limiting factor of these kernels is the memory bandwidth of the GPU, rather than the sheer compute throughput. 

We used the attainable throughput predicted by the roofline model to estimate the latency breakdown of the closed-loop dark-hole maintenance routine outlined by Pogorelyuk et al.\cite{DZM_OG} Table~\ref{tab:modeled_result} reports the modeled runtimes of the major HOWFSC kernels across four representative hardware classes: RadHard processors (BAE~RAD5545 and Teledyne~LX2160A), a radiation-tolerant edge AI accelerator (Aethero’s Nvidia Jetson Orin NX), and high-performance commercial datacenter hardware (AMD EPYC~9575F CPU and Nvidia H100~SXM GPU)\cite{BAE5545,LX2160A,orinnx,H100,9575F}.  

The results reveal a gap of approximately two orders of magnitude between RadHard platforms and modern commercial systems. They also show that EKF and EFC with precomputed matrices are similarly dominant kernels, whic is consistent with the complexity analysis in prior studies that suggest both kernels are $\mathcal{O}(\texttt{\#} \text{Channels}\times\texttt{\#} \text{Pixels}\times\texttt{\#}\text{Actuators})$ \cite{THESIS,EMBEDDED,COMPLEX}. Pairwise probing with precomputed probe influences can run much faster, as it avoids accessing the full Jacobian at every iteration. These numbers represent best-case performance: roofline predictions assume ideal resource utilization and perfect software efficiency. Real-world performance is generally worse due to memory inefficiencies, scheduling overheads, and implementation limitations.

\begin{table}[t]
\begin{center}       
\begin{tabular}{c|c|c|c|c|c} 
\hline
\rule[-1ex]{0pt}{3.5ex} \multirow{2}{*}{Kernel} & \multicolumn{5}{c}{\cellcolor{gray!20}\rule[-1ex]{0pt}{3.5ex}Estimated Runtime} \\
\cline{2-6}
\rule[-1ex]{0pt}{3.5ex} & RAD 5545*\cite{BAE5545} & LX2160A\cite{LX2160A} & Orin NX\cite{AETHERO,orinnx}\textdagger* & H100\cite{H100}* & EPYC 9575F\cite{EPYC_PERF,9575F} \\
\hline
\rule[-1ex]{0pt}{3.5ex} PWP& \SI{19.4}{\milli\second} & \SI{2.04}{\milli\second} & \SI{282}{\micro\second} & \SI{8.58}{\micro\second} & \SI{24.9}{\micro\second}\\
\hline
\rule[-1ex]{0pt}{3.5ex} EKF& \SI{16.6}{\second} & \SI{1.84}{\second} & \SI{1.04}{\second} & \SI{31.6}{\milli\second} & \SI{91.8}{\milli\second}\\
\hline
\rule[-1ex]{0pt}{3.5ex} EFC& \SI{17.4}{\second} & \SI{1.93}{\second} & \SI{1.09}{\second} & \SI{33.1}{\milli\second} & \SI{96.3}{\milli\second}\\
\hline
\end{tabular}
\end{center}
\caption{Roofline model predicted runtime of HOWFSC algorithms adapted from Eickert and Pogorelyuk et al \cite{COMPLEX,THESIS}. Machines labeled with * don't have enough physical memory to fit the precomputed matrix, and would be significantly slower in actual workload. \textdagger: The Orin NX has limited support for double precision computation, actual computation may be much slower or infeasible. } 
\label{tab:modeled_result}
\end{table} 

\subsection{Hardware Survey}

\begin{figure}[t]
\begin{center}
\begin{tabular}{c c}
\includegraphics[width=0.45\linewidth]{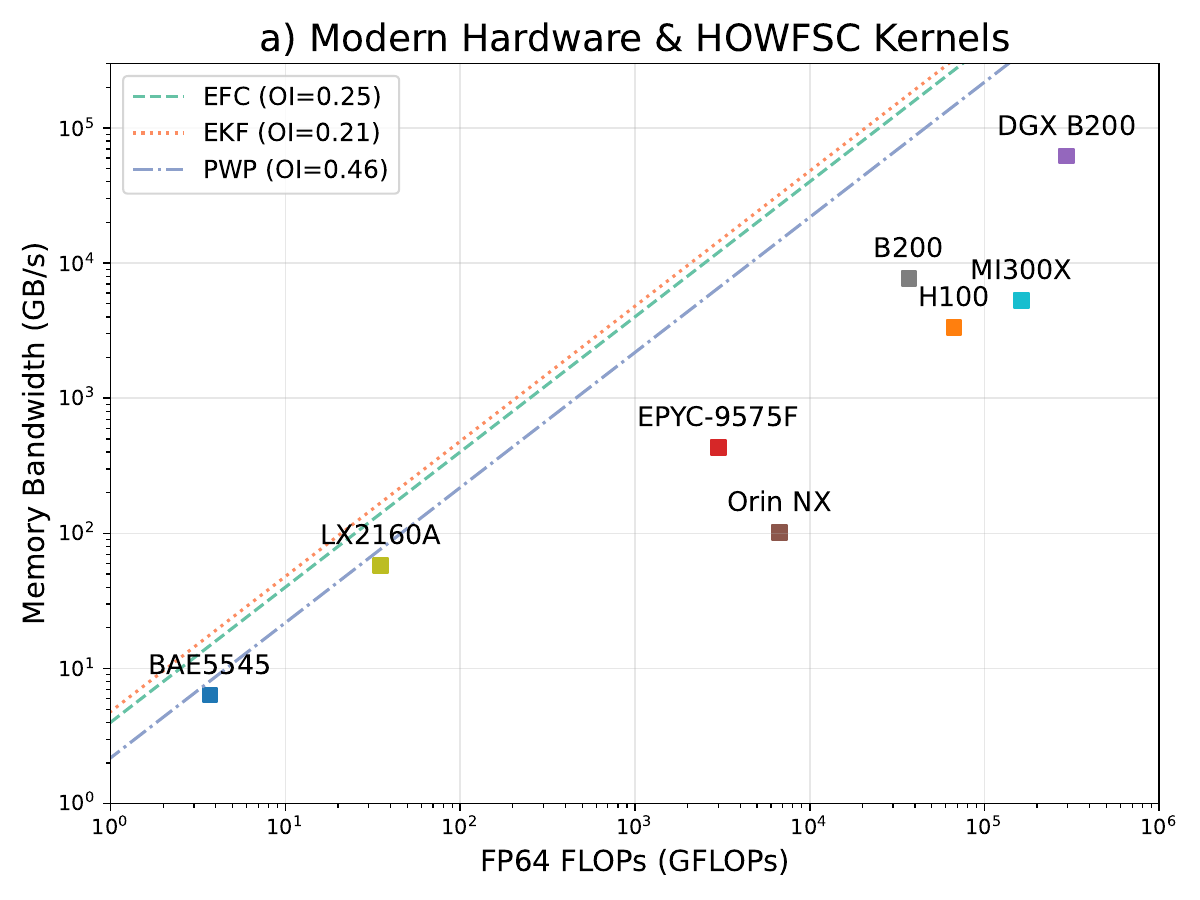} & \includegraphics[width=0.45\linewidth]{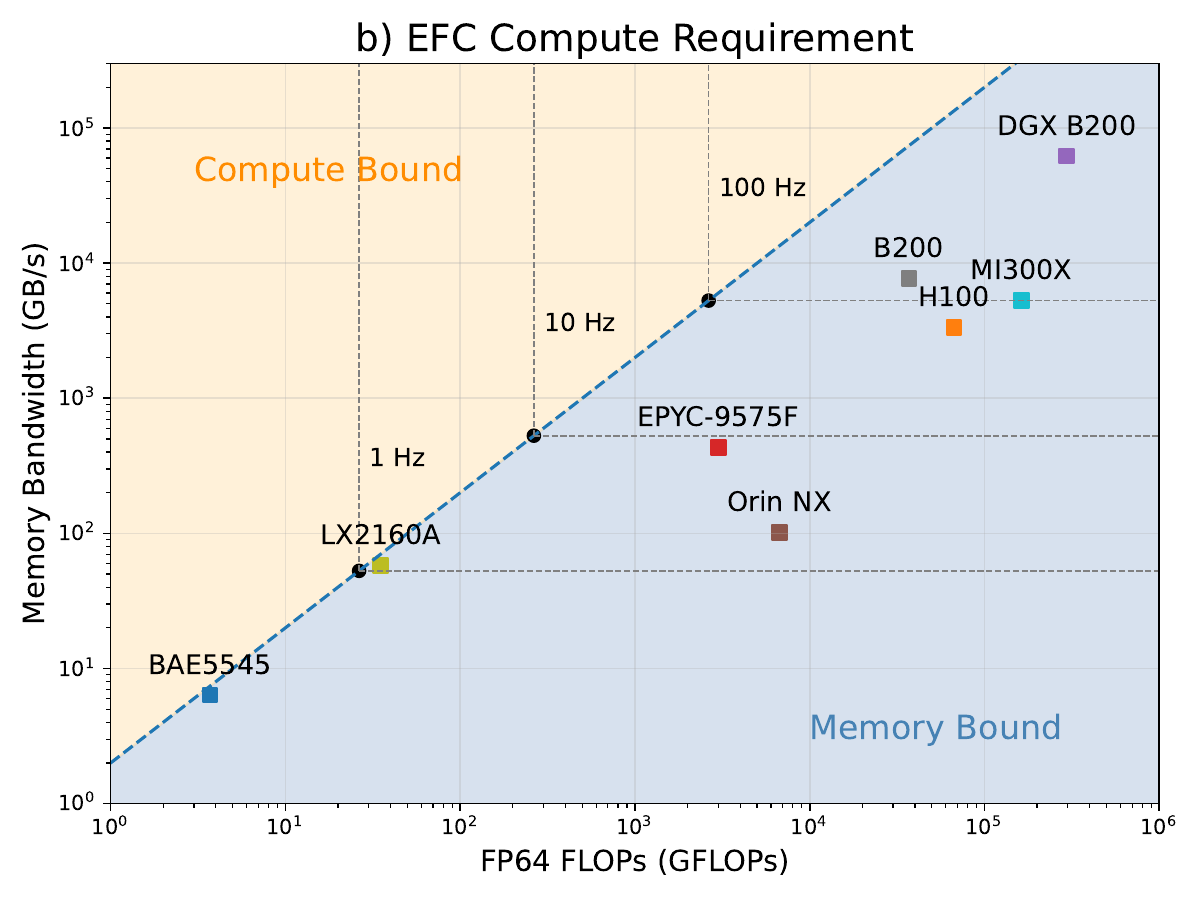}
\end{tabular}
\end{center}
\caption 
{ \label{fig:hw_landscape}
a) Modern high-performance hardware by double-precision floating-point performance and memory bandwidth, plotted with the operational intensity of HOWFSC Kernels. b) EFC compute requirement at different iterations per second (considering only EFC).} 
\end{figure} 

Figure~\ref{fig:ROOF} shows that the HOWFSC kernels are memory-bound on the H100 GPU. Applying the same roofline analysis to the remaining hardware platforms revealed that this behavior generalizes across architectures. Figure~\ref{fig:hw_landscape}a plots each system’s double-precision throughput and sustained memory bandwidth against the operational intensity of the HOWFSC kernels, where points farther to the right indicate higher FLOP capability and points higher on the axis indicate greater memory bandwidth\cite{BAE5545,LX2160A,9575F,orinnx,AETHERO,H100,EPYC_PERF,B200,MI300X}. The diagonal lines correspond to the compute–memory balance for each kernel: an architecture point lying below a kernel line indicates that the kernel is memory-bound on that system, whereas points above the line indicate compute-bounded execution. With the exception of the two RadHard processors which lack sufficient floating-point throughput, all modern high-performance architectures fall well below the kernel lines, demonstrating a consistent and architecture-independent memory bottleneck.

\subsubsection{Memory as the Limiting Factor}

We further evaluated hardware capabilities specifically against EFC, the dominant and most time-consuming kernel in both our modeling and in prior work\cite{COMPLEX,THESIS}. Figure~\ref{fig:hw_landscape} b) plots the compute requirements of EFC for 1, 10, and \SI{100}{\hertz} control rates. An architecture can sustain LUVOIR-scale EFC at \SI{100}{\hertz} only if it lies to the upper right of the \SI{100}{\hertz} point, which is achieved only by the highest-end accelerators or even clusters, such as the B200, MI300X, and DGX~B200. Although edge AI devices and high-performance CPUs provide sufficient floating-point throughput for \SI{100}{\hertz} operation, their memory bandwidth remains orders of magnitude too low. These results reinforce the central bottleneck for high-frequency HOWFSC: memory bandwidth rather than compute throughput.

Memory capacity creates an additional challenge. Storing the full Jacobian matrix, along with the precomputed gain matrix for a LUVOIR-scale dark hole in double-precision requires approximately \SI{110}{\giga\byte}\cite{COMPLEX}. Hardware that cannot hold the entire matrix in physical memory must fall back to disk or host-memory offloading, which could reduce the effective bandwidth by several orders of magnitude. Under these conditions, high-frequency EFC becomes infeasible regardless of the raw compute capability.

\subsubsection{The Hardware Trend for Lower Precision}

In our analysis, we conservatively adopted a double-precision floating point, because the effect of numerical precision on end-to-end contrast remains insufficiently characterized. This introduces an additional hardware challenge for EFC, namely the requirement for high-precision arithmetic in conjunction with extremely high memory bandwidth. Architectures that provide large memory capacity and bandwidth are increasingly optimized for AI workloads, where reduced-precision datatypes dominate, leading many commercial accelerators to offer limited or severely de-emphasized FP64 support \cite{B200,TPUv4}. This trend has been evident in recent GPU designs. For example, while the NVIDIA B300 substantially increases memory capacity and modestly improves memory bandwidth relative to the B200, its peak double-precision throughput is reduced from up to 74 TFLOPs to approximately 1.2 TFLOPs \cite{B200,B300}, reflecting a deliberate reallocation of silicon toward low-precision compute. As a result, the B300 becomes compute-bound for the EFC kernel and delivers significantly lower performance than the B200 despite its improved memory subsystem. Given the continued industry shift toward reduced-precision acceleration driven by AI workloads, reliance on commercial hardware is likely to become increasingly misaligned with the numerical requirements of HOWFSC.

At the same time, systematic studies of reduced-precision formulations for HOWFSC could substantially relax memory capacity and bandwidth demands, potentially lowering system cost and complexity. Such investigations remain an important direction for future work, but are beyond the scope of this study.

\section{Architecting Compute Systems for High Frequency HOWFSC}

Our analysis of HOWFSC algorithms and existing hardware reveals three primary design requirements for an effective compute system: high memory bandwidth, sufficient memory capacity to store large Jacobian matrices, and high double precision throughput. These requirements point toward a memory-centric system design methodology in which data movement and memory subsystem performance dominate overall capability.

Space computing introduces additional constraints, most notably energy efficiency and radiation tolerance. Power generation and thermal dissipation limits render the straightforward adoption of datacenter-class hardware impractical\cite{SPACE_DC}. Reducing energy consumption allows co-flying satellites to be smaller, lighter, and less costly, which offers clear economic advantages. Reliability presents an even greater challenge. Commercial off-the-shelf hardware, although highly performant, is not designed for high-radiation environments and may experience incorrect computation or even critical failures that jeopardize the spacecraft. 

This section focuses on architectural considerations for performance and energy efficiency, including an analysis of potential memory technologies, and of workload parallelization characteristics. A discussion of radiation challenges and protection strategies is presented in Sec.~\ref{sec:radiation}.

\subsection{The Memory Technology Spectrum}


\begin{table}[t]
\begin{center}       
\begin{tabular}{c|c|c|c|c|c} 
\hline
\rowcolor{gray!20}\rule[-1ex]{0pt}{3.5ex} Memory Type & Density& Bandwidth  & Energy & Non-Volatility & Native Radiation Resistance\\
\hline
DRAM & $\sim$ & $\sim$ & $\sim$ & \cross& \cross\\
SRAM & \reddown\reddown & \greenup\greenup\greenup & \greendown\greendown  & \cross& \cross  \\
HBM  & \greenup & \greenup & \greendown &\cross &\cross\\
NAND & \greenup\greenup\greenup & \reddown\reddown & \redup\redup & \cmark & \cross \\
ReRAM & \reddown\reddown\reddown & \greenup\greenup & $\sim$ &\cmark &\cmark \\
MRAM & \reddown\reddown\reddown & \greenup\greenup & $\sim$&\cmark &\cmark \\
\hline
\end{tabular}
\end{center}
\caption{A qualitative overview of the tradeoffs between memory technologies \cite{2NM_SRAM,RRAM,AMBER_RRAM,MRAM, HBM, HBM_PHY, DDR,NAND}.  } 
\label{tab:mem_spectrum}
\end{table} 

Since memory bandwidth and capacity are the primary limiting factors for HOWFSC, we provide a brief survey of the memory technology landscape. Across established memory classes, bandwidth and capacity exhibit a fundamental tradeoff: Static Random Access Memory (SRAM) offers the highest bandwidth but the lowest density, Dynamic Random Access Memory (DRAM) provides intermediate bandwidth and capacity, and NAND-based storage achieves extremely high density at the cost of very low bandwidth. The rapid growth of AI workloads has driven substantial innovation in memory architectures, expanding this traditional hierarchy and introducing new technologies with distinct performance, energy, and packaging characteristics. Table~\ref{tab:mem_spectrum} summarizes the resulting tradeoffs at a high level. We intentionally present these comparisons qualitatively rather than through fixed numerical values, as achievable bandwidth, capacity, and energy vary widely across vendors, process nodes, packaging options, and product generations, and point estimates are often not representative of the broader design space or long-term technology trends.

\paragraph{High Bandwidth Memory (HBM)}
HBM has become the dominant memory technology in modern AI accelerators due to its ability to provide very high bandwidth\cite{TPUv4,B200,MI300X}. It achieves this by stacking DRAM dies and integrating them closely with the processor using advanced packaging techniques, which dramatically increases throughput compared to conventional double data rate (DDR) based DRAM. As shown in Fig. \ref{fig:hw_landscape}, the four architectures with the highest sustained memory bandwidth all employ HBM, forming a clear separation from systems that rely on DDR-DRAM.

\paragraph{Resistive RAM (ReRAM) and Magnetoresistive RAM (MRAM)}
ReRAM stores information by modulating the resistive state of a memory element. It offers several attractive properties, including high radiation tolerance, low read energy, and high bandwidth, but its density currently remains limited. MRAM stores data using magnetic domains and shares many of the same advantages as ReRAM, including robustness under radiation. However, MRAM also struggles to scale to densities comparable to DRAM. Continued development in these emerging memory technologies will be necessary before they can compete directly with SRAM and DRAM for large-scale deployments.


\subsection{Parallel HOWFSC System}
\label{subsec:parallel}
\begin{figure}[t]
\begin{center}
\begin{tabular}{c}
\includegraphics[width=0.6\linewidth]{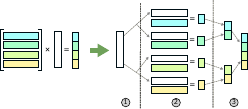}
\end{tabular}
\end{center}
\caption 
{ \label{fig:parallelEFC}
GEMV Parallelization Strategy with constant matrix: 1) Broadcast input vector states to all processors, 2) perform inner product individually, and 3) gather output vector. } 
\end{figure}

HOWFSC computations are naturally amenable to parallelization. The EFC kernel consists of two consecutive GEMV operations, first between $\mathbf{J}^T$ and $\mathbf{E}$, and secondly between the gain matrix $M$ and their product. GEMV can be efficiently partitioned across multiple processing elements. To distribute GEMV, the matrix is divided into chunks, and each processing element independently computes its assigned portion of the matrix–vector product. The partial results are then gathered by a leader processing element, as illustrated in Fig. \ref{fig:parallelEFC}. With $N$ processors, the compute throughput, memory bandwidth, and memory capacity required per processor all decrease by a factor of $N$.

The corresponding communication overhead is also low. Before computation, the electric field state vector is broadcast to all processors. Although the broadcast appears to scale with $N$, modern interconnects implement broadcast efficiently, so the practical overhead is much smaller. After computation, each processor transmits only its $\frac{1}{N}$ segment of the control vector back to the leader.

The broadcast and gather operations can be efficiently implemented with a hierarchical-tree (H-Tree) network. As a worst-case example, consider a LUVOIR-scale coronagraph operating at \SI{100}{\hertz} (consistent with the requirements in Fig.~\ref{fig:howfsc_motiv}), using $N = 128$ processing elements connected by a 2-degree H-tree. Under these conditions, the required point-to-point bandwidth is only $\sim$ \SI{30.6}{\giga\byte\per\second} (derivation shown in Appendix \ref{sec:itcbw}), far below what modern interconnect technologies provide~\cite{INTERCON}. Increasing the degree of the tree would further reduce the bandwidth requirement, at the cost of increased per-processor complexity. These results indicate that interconnect bandwidth is not a limiting factor for parallelizing EFC.

This low communication overhead makes parallel systems a promising approach for scaling HOWFSC performance. The primary design questions then become selecting an appropriate parallelization degree and identifying suitable compute building blocks.

The GEMV operation for precomputing $\mathbf J \Delta\mathbf u$ in EKF for the subsequent HOWFSC iteration can be parallelized in the same fashion, and exhibits a low communication overhead as well.

Other HOWFSC kernels, such as the state updates in EKF and pairwise probing, are much smaller than EFC and can be executed efficiently on RadHard processors such as the 16-core Teledyne LX2160 Space, as suggested by Tab.~\ref{tab:modeled_result}.

\subsection{HOWFSC Pipeline}
\begin{figure}[t]
\begin{center}
\begin{tabular}{c}
\includegraphics[width=0.95\linewidth]{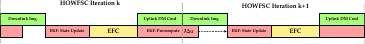}
\end{tabular}
\end{center}
\caption 
{ \label{fig:HOW_pip}
Iterative Operation of HOWFSC, overlapping the computation of $\mathbf J\Delta\mathbf u$ and inter-satellite communication. The top row represents inter-satellite link usage and the bottom row represents compute usage. (Box lengths are schematic and not to scale.)} 
\end{figure}

We model a HOWFSC dark hole maintenance pipeline outlined by Pogorelyuk et al \cite{DZM_OG}. We also include transfer overhead, considering a co-flying satellite \SI{100}{\kilo\meter} apart from the telescope, and assuming a spatial link with \SI{1}{\milli\second} latency and \SI{25}{\giga\bit\per\second} data rate between the satellite and telescope \cite{STARLINK_MINILSR}. We calculate the latency for a single HOWFSC loop, which includes: 1) Transfer captured image from telescope to co-flying satellite, 2) Convert pixels to electric field states via EKF, 3) Solve for controls with EFC, assuming precomputed matrices. 4) Add generated dither to the control vector, and lastly, 5) Transfer solved controls to the telescope. Figure \ref{fig:HOW_pip} illustrates such a HOWFSC routine.

\subsection{GPU-Based System}
We first analyze a hardware design that is based on commercial off-the-shelf hardware with minimal design changes. For such a system, we elect GPUs due to their high compute throughput and high memory bandwidth. The most advanced GPU is the Nvidia B200, providing \SI{180}{\giga\byte} HBM3e memory at \SI{7.7}{\tera\byte\per\second} bandwidth with up to 37 TFLOPs of compute throughput\cite{B200}. While it is currently only available as 8-GPU servers, a dedicated 1-GPU or 2-GPU Unit is highly plausible and would require minimal engineering efforts. 

\begin{figure}[t]
\begin{center}
\begin{tabular}{c}
\includegraphics[width=0.95\linewidth]{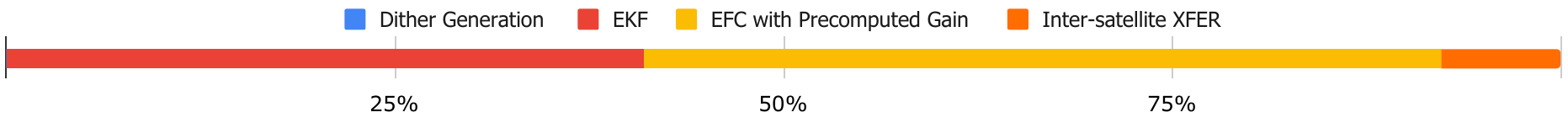}
\end{tabular}
\end{center}
\caption 
{ \label{fig:B200}
Latency breakdown of a dark hole maintenance iteration on Nvidia B200.} 
\end{figure}

On a single B200, our roofline models suggest a total latency of \SI{28.1}{\milli\second}, or a control frequency of \SI{35.5}{\hertz}. Figure \ref{fig:B200} shows the runtime breakdown of the different components. EFC and EKF are the dominant portion of the pipeline by far, with similar runtimes. Appendix \ref{sec:validation} shows our validation of the EFC kernel execution.

\begin{figure}[t]
\begin{center}
\begin{tabular}{c}
\includegraphics[width=0.95\linewidth]{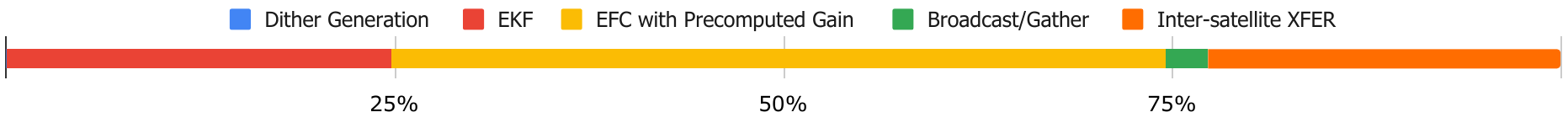}
\end{tabular}
\end{center}
\caption 
{ \label{fig:3B200}
Latency breakdown of a dark hole maintenance iteration on a three B200 GPU cluster.} 
\end{figure}

Using the parallelization strategy described in
Sec.~\ref{subsec:parallel}, the control loop can be scaled to higher frequencies by distributing computation across multiple GPUs. For a cluster of three B200 GPUs, the roofline model predicts a total iteration latency of \SI{9.65}{\milli\second}, corresponding to a maximum control frequency of \SI{103.7}{\hertz}. Figure~\ref{fig:3B200} shows the resulting latency breakdown. As compute time decreases with additional GPUs, inter-satellite data transfer constitutes an increasing fraction of the iteration latency. This shift enables a larger portion of EKF precomputation to be overlapped with communication, further improving end-to-end efficiency.

\subsection{Inefficiencies in GPU}
GPUs are optimized for throughput rather than predictable execution, leading to non-deterministic kernel latency arising from dynamic warp scheduling, shared memory hierarchy contention, and system-level interference. Their energy consumptions are extremely high: each B200 consumes up to a kilowatt of power, and hosting three B200s that can consume a combined \SI{3}{\kilo\watt} poses a tremendous challenge on the power generation and heat dissipation capabilities of the satellite (Sec. \ref{sec:cost}).

We recognize that executing HOWFSC on GPU is extremely inefficient, as there is a significant mismatch between feature that GPU provides compared to the HOWFSC requirements.

\paragraph{Compute Mismatch}
HOWFSC kernels exhibit low operational intensities, but GPUs are optimized for much higher operating intensity (OI), and a significant amount of arithmetic units are idle during the computation. The latest GPUs are also optimized more for low-precision AI workloads, and much of the computing units stay idle.

\paragraph{Redundant Caching}
GPUs have multiple layers of on-chip SRAM caches to effectively capture data reuse and improve memory access bandwidth and latency. However, the Jacobian size for EFC far exceeds the capacities of these on-chip SRAM caches, and EFC's GEMV workload exhibits no data reuse. These SRAM caches are thus irrelevant to the performance of EFC, and moving data across the memory hierarchy only results in wasted energy.

\section{Energy Efficient HOWFSC Accelerators}

GPUs are poorly matched to HOWFSC from an energy-efficiency standpoint, and computational satellite cost modeling shows that compute energy consumption is the dominant contributor to overall satellite cost~\cite{bleier2025architecting}. This dominance arises from the need to provision for worst-case power and energy requirements, as well as the cascading overheads such provisioning imposes on other satellite subsystems. Together, these factors motivate the development of high-level architectures tailored to the characteristics of HOWFSC, with the goal of substantially improving energy efficiency without compromising performance. We therefore explore two design strategies: a lower-cost solution based on HBM and a higher-cost architecture built around distributed SRAM.

\begin{table}[t]
\centering
\begin{tabularx}{\columnwidth}{l|YYYYY}
\hline
\multirow{2}{*}{\textbf{HBM}\cite{HBM_PHY, HBM}} 
  &  \cellcolor{gray!20} Capacity & \cellcolor{gray!20} Area/Chip & \cellcolor{gray!20} Throughput & \cellcolor{gray!20} PHY Area & \cellcolor{gray!20} Read Energy \\
\cline{2-6}
  & \rule[-1ex]{0pt}{3.5ex} \SI{48}{\giga\byte} 
  & \SI{121}{\milli\meter\squared} 
  & \SI{1.125}{\tera\byte\per\second} 
  & \SI{8.83}{\milli\meter\squared} 
  & \SI{3.69}{\pico\joule\per\bit} \\
\hline
\multirow{2}{*}{\textbf{SRAM}\cite{2NM_SRAM, SRAM_LEAK}} 
  & \cellcolor{gray!20} Tech Node & \cellcolor{gray!20} Density & \cellcolor{gray!20} Max Freq. & \cellcolor{gray!20} Read Energy & \cellcolor{gray!20} Leakage \\
\cline{2-6}
  & \rule[-1ex]{0pt}{3.5ex} TSMC 2nm 
  & \SI{38.1}{\mega\bit\per\milli\meter\squared} 
  & \SI{3}{\giga\hertz} 
  & \SI{140.8}{\femto\joule\per\bit} 
  & \SI{18.75}{\pico\watt\per\bit} \\
\hline
\multirow{2}{*}{\textbf{EFC Compute}\cite{ASAP7}} 
  & \cellcolor{gray!20} Tech Node & \cellcolor{gray!20} Macro & \cellcolor{gray!20} Freq. & \cellcolor{gray!20} Area/Macro & \cellcolor{gray!20} Power/Macro \\
\cline{2-6}
  & \rule[-1ex]{0pt}{3.5ex} ASAP 7nm 
  & FP64 MAC 
  & \SI{500}{\mega\hertz} 
  & \SI{1450.9}{\micro\meter\squared} 
  & \SI{1.49}{\milli\watt} \\
\hline
\multirow{2}{*}{\textbf{Interconnect}\cite{PARALLELIO}} 
  & \cellcolor{gray!20} Tech Node & \cellcolor{gray!20} Width & \cellcolor{gray!20} Bandwidth & \cellcolor{gray!20} Area/Macro & \cellcolor{gray!20} Energy \\
\cline{2-6}
  & \rule[-1ex]{0pt}{3.5ex} 32nm 
  & 16-Lane 
  & \SI{16}{\giga\bit\per\second} 
  & \SI{1.3}{\milli\meter\squared} 
  & \SI{2.6}{\pico\joule\per\bit}\\
\hline
\multirow{2}{*}{\textbf{GP Compute}\cite{LX2160A}} 
  & \cellcolor{gray!20} Model & \cellcolor{gray!20} Throughput  & \cellcolor{gray!20} Mem BW & \cellcolor{gray!20} Area/Chip & \cellcolor{gray!20} Power \\
\cline{2-6}
  & \rule[-1ex]{0pt}{3.5ex} LX2160A
  & \SI{35.2}{\giga Flop\per\second}
  & \SI{57.6}{\giga\byte\per\second} 
  & \SI{1600}{\milli\meter\squared} 
  & \SI{25}{\watt}\\
\hline
\end{tabularx}
\vspace{0.5em}
\caption{Parameters Used for Custom Architecture Exploration. 
For SRAM configuration, since TSMC did not release 2nm SRAM energy 
numbers, we use conservative energy metrics from a 22nm SRAM that is 
optimized for leakage. For compute configuration, we obtain the area, 
power, and frequency metrics by synthesizing the Synopsys DesignWare 
double precision multiply accumulate unit (MAC) with the ASAP7 
predictive PDK, which offers conservative approximates for TSMC 2nm.}
\label{tab:parameters}
\end{table}

Table \ref{tab:parameters} shows the parameters used in our architectural exploration, based on available information on state-of-the-art technology, with some unavailable metrics conservatively derived from older technologies. Considering that the HWO is planned for 2040, we assume that the custom processor will be fabricated in a process node with equivalent or greater performance than TSMC 2nm, and that DRAM solutions are equivalent or better than HBM3E,  leaving room for technological advancement.

\begin{table*}[t]
\centering
\begin{minipage}[t]{0.45\textwidth}
\centering
\begin{tabular}{l|c}
\hline
\multicolumn{2}{c}{\cellcolor{gray!20}\rule[-1ex]{0pt}{3.5ex}\textbf{HBM-Based Architecture}}\\
\hline
Num HBM Modules  & 27 \\
HBM PHY Power & \SI{83.3}{\watt}\\
HBM Power & \SI{605}{\watt}\\
Num Compute Macros & 7400\\
Total Area & \SI{5151}{\milli\meter\squared}\\
Total Power & \SI{928}{\watt}\\
\hline
\end{tabular}
\end{minipage}
\begin{minipage}[t]{0.5\textwidth}

\begin{tabular}{c}
\includegraphics[width=\linewidth]{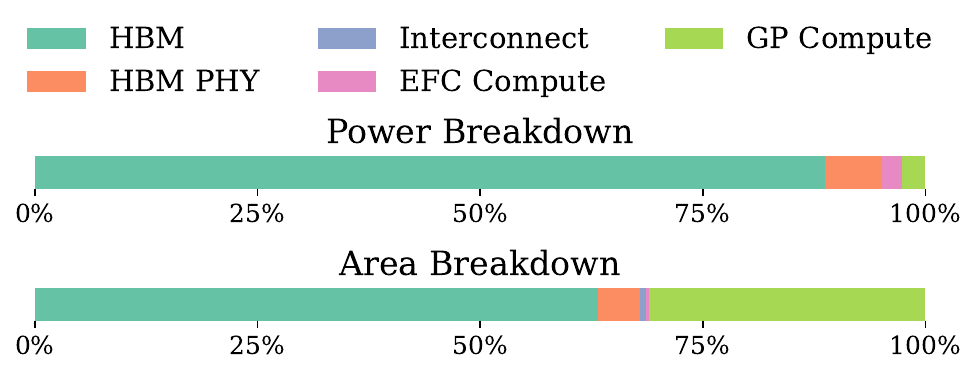}
\end{tabular}
\end{minipage}

\vspace{0.5em}

\caption{High-Level Specification for HBM system with visualizations, along with energy and area breakdowns. Reading from HBM is the predominant power consumption of the system. HBM and GP compute packages takes up most of the area. }
\label{tab:HBMSys}
\end{table*}

\subsection{HBM-based Process-Near-Memory Architecture}
Processing near memory architectures move compute closer to the memory; they are optimized for low operational intensity,  memory-bound workloads. Targeting \SI{100}{\hertz} for an iso-control frequency comparison against GPU, we propose a high-level HBM-based architecture employing multiple HBM stacks to offer high aggregated memory bandwidth and capacity. The compute chip will be packed with double-precision multiply-accumulate units that compute the matrix vector multiplication in parallel.  Table \ref{tab:HBMSys} shows an example HBM system specification: 27 HBM modules together provide enough bandwidth to sustain the computation, more than 7000 MAC units offer the compute throughput required for EFC at \SI{100}{\hertz} for LUVOIR A. The precomputed matrix will be equally partitioned across the 27 HBM modules, and the compute elements would fetch these stored entries consecutively, and divide up work between the individual processing elements. While the exact layout, such as how much compute needs to be paired with how many HBM modules, depends on physical constraints on HBM placement and chip dimensions and has implications on the precise performance and energy consumption, we factor out such details for a high-level analysis.

To simplify the system design, only EFC and the EKF precomputation $\mathbf J\Delta\mathbf u$ will be carried out in custom HBM architecture, and we employ a 16-core Radhard Teledyne LX2160 Space processor for the rest of the kernels, including per-pixel EKF state updates and dither generation/addition. This approach would significantly reduce power consumption from \SI{3}{\kilo\watt} to \SI{939}{\watt} while achieving the same HOWFSC performance.


The downside of HBM is that it has low radiation tolerance and is very inflexible. HBM options are limited and adding additional radiation tolerance using methods as described in Sec. \ref{sec:radiation} is highly impractical.

\subsection{SRAM-Only Distributed Memory Architecture}

\begin{figure}[t]
\begin{center}
\begin{tabular}{c}
\includegraphics[width=0.7\linewidth]{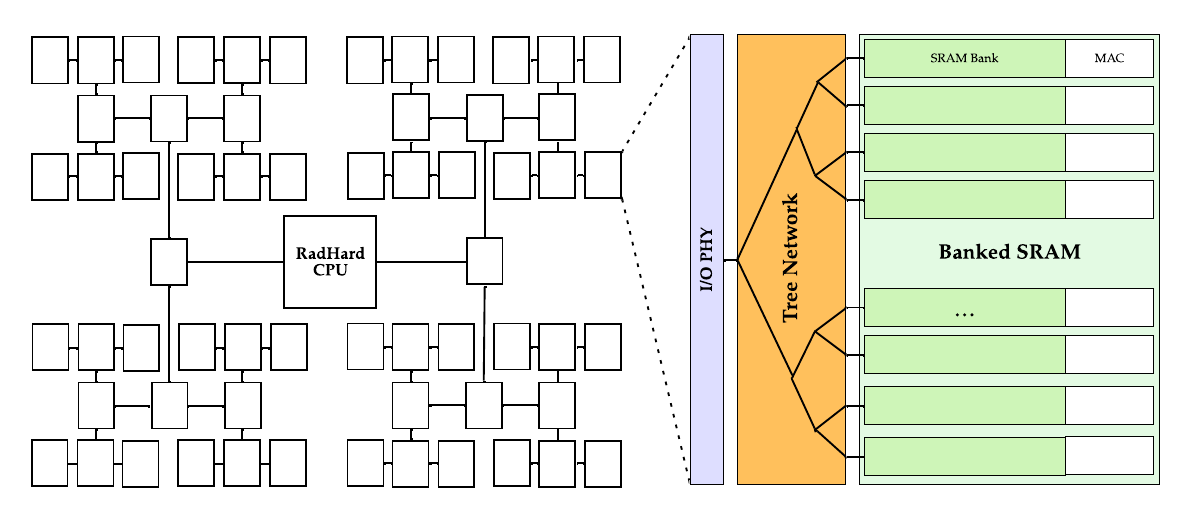}
\end{tabular}
\end{center}
\caption 
{ \label{fig:SRAM}
High-level architecture of an SRAM-only distributed memory system. 62 SRAM-chips are arranged in a 2-degree 5-tier hierarchical tree interconnect. Input electric field state vectors are broadcast through an off-chip interconnect first, and then distributed the SRAM banks through an on-chip tree network. Each SRAM bank is co-located with multiply-accumulate compute elements to compute local dot products. Results are then gathered to the CPU, traversing the network hierarchy in reverse direction.} 
\end{figure} 

SRAMs, on the other hand, are highly customizable, allowing for radiation tolerant designs. The downside of SRAM is mainly its low density compared to DRAM or NAND, with the latest SRAM capacity only at \SI{38}{\mega\bit\per\milli\meter\squared} \cite{2NM_SRAM}, to store the full Jacobian requires more than \SI{22000}{\milli\meter\squared} of SRAM, far exceeding the area limit of a single chip. It is also less economical, as more silicon need to be fabricated in a more advanced technology node, hence large capacity SRAM-based chips have been inviable for mass production consumer electronics. However, in the case of HOWFSC, these hardware costs pale in comparison to the cost of the satellite and  operational costs of managing a satellite at the Sun-Earth L2 Lagrange point.

With a SRAM-only distributed memory architecture, we co-locate compute elements close to SRAM macros, and distribute the precomputed Gain matrix to SRAM banks. We allocate compute elements to SRAM macros by their capacity, designing a memory centric system where the compute capability precisely matches the requirements of EFC. Table \ref{tab:SRAMSys} shows example design parameters for such a system with equal performance compared to the HBM and GPU-based solutions. Similar to the HBM design, we let the Radhard CPU handle kernels other than EFC.

\begin{figure}[t]
\begin{center}
\begin{tabular}{c}
\includegraphics[width=0.95\linewidth]{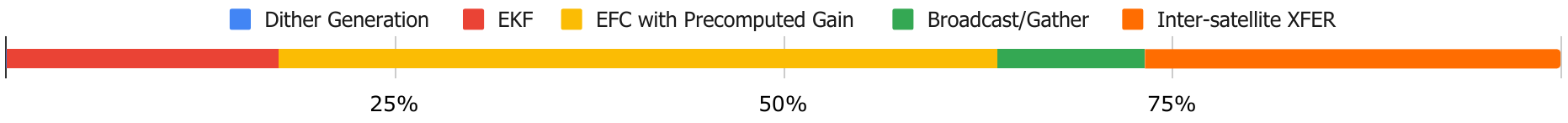}
\end{tabular}
\end{center}
\caption 
{ 
Latency breakdown of the SRAM Solution.} 
\label{fig:SRAM_breakdown}
\end{figure}

This system offers tremendous energy savings with just \SI{84.65}{\watt} of operating power at the cost of increased hardware costs and increased system physical dimension. Figure \ref{fig:SRAM} shows a high-level system architecture diagram.

\begin{table*}[t]
\centering
\begin{minipage}[t]{0.45\textwidth}
\centering
\begin{tabular}{l|c} 
\hline
\multicolumn{2}{c}{\cellcolor{gray!20}\rule[-1ex]{0pt}{3.5ex}\textbf{SRAM-Only Architecture} }\\
\hline
Num SRAM Chiplets & 56 \\
\hline
MAC Unit Per-Chiplet & 135\\
\hline
SRAM Dynamic Power & \SI{38.2}{\watt} \\
\hline
SRAM Leakage & \SI{10.9}{\watt} \\
\hline
Total Area & \SI{23879}{\milli\meter\squared}\\
\hline
Total Power & \SI{84.65}{\watt}\\
\hline
\end{tabular}
\end{minipage}
\begin{minipage}[t]{0.50\textwidth}

\begin{tabular}{c}
\includegraphics[width=\linewidth]{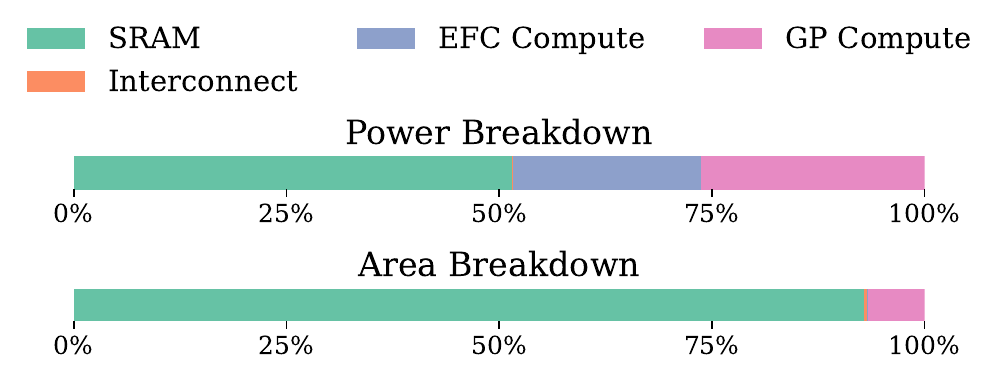}
\end{tabular}
\end{minipage}

\vspace{0.7em}

\caption{High-Level Specification for SRAM system. SRAM consumes significantly less energy compared to HBM, however, it incurs an enormous area overhead.} 
\label{tab:SRAMSys}
\end{table*}

\subsection{Power Scaling}

\begin{figure}[t]
\begin{center}
\begin{tabular}{c}
\includegraphics[width=0.55\linewidth]{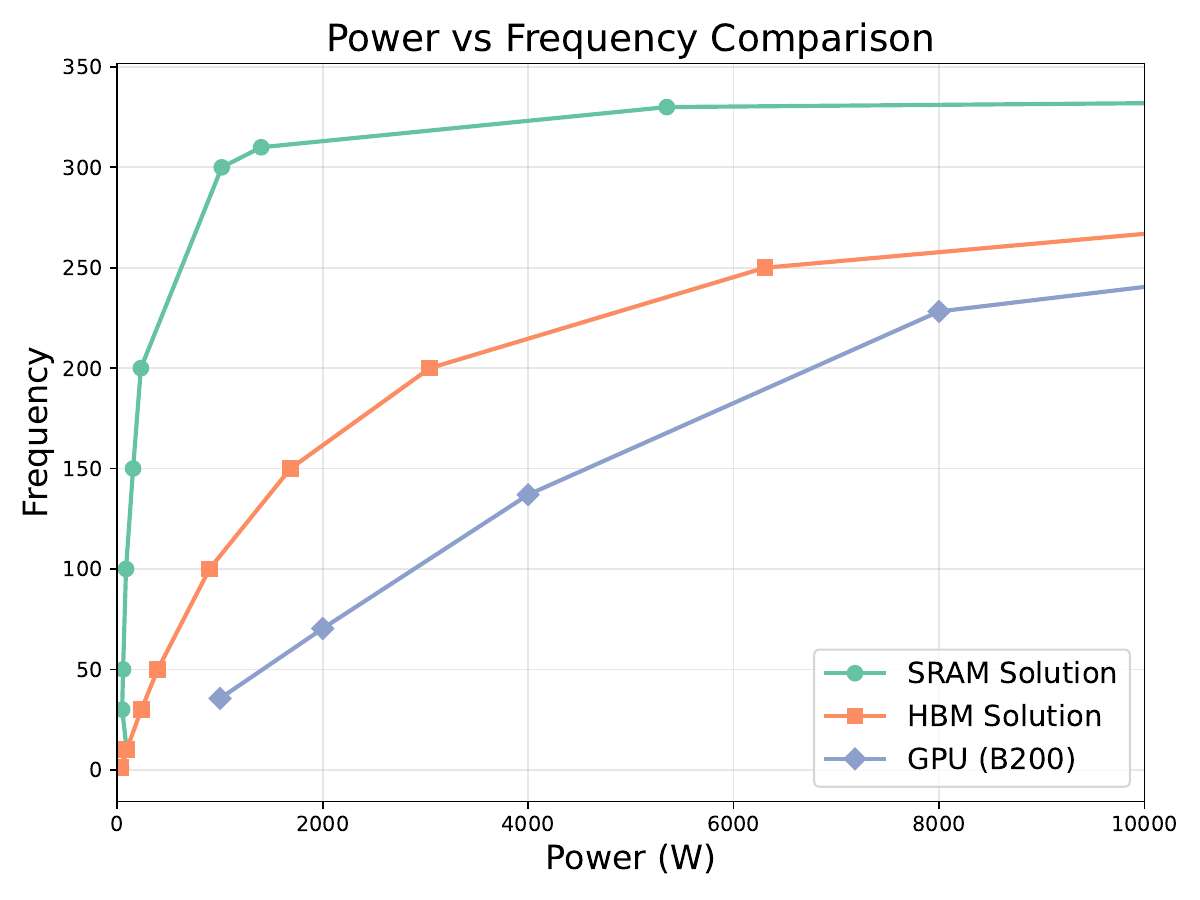}
\end{tabular}
\end{center}
\caption 
{ \label{fig:pvf}
Power vs. Frequency for the three hardware solutions}
\end{figure} 

We analyze the scalability and power efficiency of the three candidate compute systems for HOWFSC by sweeping multiple design points across a broad range of target control frequencies. Figure \ref{fig:pvf} plots these design points and illustrates how each architecture scales as system power increases. We first observe that all three systems eventually plateau in achievable frequency due to communication and data-transfer bottlenecks, rather than insufficient arithmetic throughput. However, the SRAM-based and HBM-based custom designs reach this plateau at far lower power levels than the GPU baseline. In both custom architectures, eliminating or reducing off-chip DRAM traffic allows the system to saturate performance with only a few hundred watts, while the GPU requires well over a kilowatt before approaching the same frequency ceiling. This comparison highlights that only memory-centric architectures can efficiently scale HOWFSC performance within the practical power envelope of a co-flying compute satellite.

\subsection{Trade-off Analysis}

We also analyze the architectural trade-offs between the different compute designs. Consistent with the memory technology spectrum summarized in Tab.~\ref{tab:mem_spectrum}, the HBM-based solution is fundamentally bandwidth-limited, the SRAM-based solution is capacity-limited, and the GPU-based solution is constrained by architectural mismatch rather than raw compute capability. Although GPUs provide both high arithmetic throughput and high peak memory bandwidth, their memory hierarchy and execution model are optimized for high operational intensity and data reuse, which are absent in HOWFSC workloads.

The SRAM-based system must provision sufficient on-chip storage to hold the full set of precomputed matrices, requiring a very large SRAM area. This large footprint incurs substantial static power due to leakage, making the SRAM design inefficient at very low control frequencies where dynamic activity is minimal. In this low-frequency regime, the HBM-based solution is more energy efficient, since it can operate with only a small number of HBM stacks at low aggregate read bandwidth. However, this advantage rapidly diminishes as higher control frequencies are targeted. Meeting increasing bandwidth demands forces the HBM-based system to scale the number of HBM modules aggressively, leading to a steep rise in power consumption. In contrast, the SRAM-based design does not require additional memory capacity or area as frequency increases, and its power scales primarily with dynamic activity rather than memory provisioning.

The GPU-based solution exhibits a different scaling behavior. At low frequencies, GPUs incur high baseline power due to underutilized compute units, deep cache hierarchies, and memory systems designed to exploit data locality that HOWFSC does not provide. As frequency increases, GPUs can achieve high throughput, but only at the cost of operating near their peak power envelope, often exceeding a kilowatt. Unlike the custom SRAM and HBM architectures, GPUs cannot efficiently trade power for frequency, as much of their energy consumption is spent on architectural features that do not contribute to HOWFSC performance. As a result, GPUs occupy the least favorable region of the power–frequency design space despite their raw performance advantage.

\subsection{Future Outlook}

The three compute organizations evaluated in this work are intentionally anchored to memory and interconnect technologies that are either commercially available today or on a near-term product roadmap. As a result, our design points should be interpreted as conservative baselines rather than hard limits. Looking forward, the dominant opportunities for reducing system scale are in the memory subsystem. Higher-capacity HBM generations, denser 3D-stacked SRAM or embedded memory, and improved memory energy efficiency would directly relax both the capacity and bandwidth constraints that set the number of modules (HBM) or total silicon area (SRAM) required at a given control frequency. Because our highest-frequency points are bandwidth-bound while our lowest-frequency points are often dominated by leakage and provisioning overheads, improvements in pJ/bit and capacity per package can translate into disproportionate reductions in required modules, board area, and power.

A second path is improved compute and memory locality. Emerging compute-near-memory and processing-in-memory approaches could further reduce data movement by fusing parts of the EFC and GEMV pipeline into the memory stack, thereby lowering effective bytes transferred per iteration even when arithmetic intensity is fixed. Such approaches are especially relevant for the matrix and vector workloads that dominate HOWFSC, where performance and energy are frequently limited by memory traffic rather than peak FLOPs.

Finally, higher control frequencies depend not only on raw throughput but also on interconnect efficiency and system integration. Continued reductions in energy per bit and increases in link bandwidth, across board-level, chiplet, and optical spatial links, would decrease the power and mass penalties of distributing memory and compute across multiple modules or companion spacecraft. Together, these trends suggest a clear scaling trajectory. As memory capacity and energy efficiency improve and interconnect costs fall, future systems can achieve higher cadence with fewer modules, lower power, and reduced thermal disturbance to the observatory.

\section{System-Level Cost Analysis}
\label{sec:cost}

We further analyze the costs associated with deploying a high-performance co-flying compute satellite, including both the onboard compute subsystem and the satellite platform itself. Satellite costs are modeled using an extended version of the Small Satellite Cost Model (SSCM) \cite{bleier2025architecting}, which estimates spacecraft mass, power subsystem sizing, and total cost as a function of beginning-of-life (BOL) power and payload requirements.

We evaluate three representative satellite design points corresponding to the three compute architectures analyzed in Sec.~5: the SRAM-based custom accelerator, the HBM-based custom system, and a GPU-based solution. All satellites are assumed to have a three-year operational lifetime and a 25~Gbps inter-satellite link (ISL) to support high-rate data exchange with the observatory. Table~\ref{tab:satellite_cost} summarizes the resulting estimates for BOL power, spacecraft mass, and total cost, broken down into non-recurring engineering (NRE) costs (development, integration, and testing) and recurring unit (RE) costs (manufacturing and launch).

Within the SSCM framework, spacecraft cost is driven primarily by power provisioning and mass, with secondary contributions from avionics, structural margins, and thermal control. Across the design points considered here, power and mass dominate overall cost, while other subsystems contribute only modest variation.

\begin{table}[t]
\centering
\small
\begin{tabular}{c c c c c c}
\hline
\rowcolor{gray!20}
\textbf{Compute Power} &
\textbf{BOL Power} &
\textbf{Mass} &
\textbf{Total Cost} &
\textbf{NRE} &
\textbf{RE} \\
\hline
\SI{90}{\watt}   & \SI{237}{\watt}  &  \SI{231}{\kilo\gram} & \$37.54M & \$16.7M & \$20.7M \\
\SI{900}{\watt}  & \SI{1400}{\watt}  & \SI{476}{\kilo\gram} & \$65.36M & \$29.0M & \$36.4M \\
\SI{3000}{\watt} & \SI{4418}{\watt} & \SI{1110}{\kilo\gram} & \$110.77M & \$49.4M & \$61.4M \\
\hline
\end{tabular}
\vspace{1em}
\caption{Satellite NRE and RE estimates for three different compute power provisions. }
\label{tab:satellite_cost}
\end{table}

Assuming a 25-year operational lifetime for HWO, each co-flying compute satellite must be replaced multiple times over the mission duration. Under the three-year satellite lifetime assumption, this corresponds to nine satellite units over the mission. Using the cost estimates in Table~\ref{tab:satellite_cost}, the resulting total mission cost for the satellite (excluding the onboard compute) is approximately \$203M for the 90~W design point, \$357M for the 900~W design point, and \$602M for the 3000~W design point, including both non-recurring engineering and recurring unit manufacturing/launch costs. These results highlight the strong dependence of long-term mission cost on onboard compute power, with multi-kilowatt-class systems nearly tripling the total cost relative to low-power alternatives.

We further consider the development costs of the compute hardware itself and its implications on satellite design. Direct procurement of commercial GPUs incurs minimal non-recurring cost, with negligible unit prices compared to spacecraft power provisioning, mass, and launch costs. However, such platforms drive the system toward kilowatt-class power budgets, substantially increasing spacecraft mass and total mission cost.

A custom compute system offers tradeoff by significantly reducing satellite power and mass, but it requires the fabrication of dedicated silicon. Unlike complex general-purpose processors, whose non-recurring engineering (NRE) costs are dominated by extensive software development, validation, and verification effort \cite{UCIE_COST}, the proposed HBM and SRAM-based systems designs are architecturally simple. They consist primarily of memory interfaces and dense floating-point datapaths, with minimal control logic and little reliance on a sophisticated software stack. As a result, verification and software development costs are substantially reduced.

Consequently, we expect non-recurring costs to be dominated by photomask fabrication and IP licensing rather than design or verification effort. While detailed photomask pricing is not publicly disclosed by foundries, academic studies of advanced lithography and photomask economics consistently indicate that a full mask set at leading-edge process nodes carries a cost on the order of tens of millions of dollars, with representative estimates of approximately \$30M for a 5 nm-class process \cite{HW_NEURON,UCIE_COST}. Recurring per-chip manufacturing cost is therefore expected to be negligible relative to overall satellite system cost.

Overall, this analysis shows that satellite cost scales sharply with onboard compute power, and that improvements in compute energy efficiency yield disproportionate system-level benefits when amortized over a multi-decade mission lifetime. Although absolute cost depends on assumptions such as satellite lifetime and launch pricing, the relative ordering of architectures remains unchanged: low-power, memory-centric compute systems substantially reduce total mission cost compared to kilowatt-class alternatives. These results reinforce the importance of architectures that meet HOWFSC performance targets within modest power envelopes, where both satellite and compute development costs remain tractable.

\section{Protecting Non-Radiation-Hardened Systems in Space}
\label{sec:radiation}

A key concern for deploying high-performance hardware in a co-flying satellite
is radiation tolerance. Unlike RadHard components
fabricated in specialized processes, hardware built on standard CMOS processes
offers superior performance but lacks explicit radiation hardening. However,
such hardware can be protected through a combination of inherent material
properties, system architecture, and software techniques. This section examines
the radiation environment at L2, the evolution of radiation tolerance in modern
standard CMOS technologies, and mitigation strategies suitable for HOWFSC
compute systems.

\subsection{Radiation Environment at L2}
The Sun-Earth Lagrange Point 2 (L2) presents a distinct radiation environment from low Earth
orbit (LEO). At L2, spacecraft operate outside Earth's protective magnetosphere
and are exposed to three primary radiation sources: \glspl{gcr}, \glspl{spe},
and solar wind electrons\cite{JWST_RAD, L2_RAD}.

Compared to low Earth orbit, L2 presents a moderate radiation environment:
spacecraft operate outside Earth's trapped radiation belts, with \gls{tid}
dominated by \glspl{spe} and \glspl{gcr}. Studies for JWST estimated
approximately \SI{2.4}{\kilo\rad(Si)\per year} behind 100 mils of aluminum
shielding\cite{NGST_RAD}. A key advantage of the co-flying architecture is that
the compute satellite can be designed as a replaceable unit, decoupling its
lifetime from the multi-decade HWO mission and relaxing \gls{tid} requirements.
Nevertheless, the system must still tolerate \glspl{see} from individual
energetic particle strikes during operation.


\subsection{Design Considerations for Co-flying HOWFSC}

Radiation tolerance in co-flying compute platforms must be addressed holistically across compute, memory, and communication components rather than at the device level alone. While advanced process technologies reduce the sensitive volume of logic, transient faults and data corruption remain a concern for both compute and memory subsystems in space environments. We therefore adopt a defense-in-depth approach that combines architectural redundancy, error detection and correction, and system-level recovery mechanisms. Appendix \ref{sec:rad_appx} discusses the increasing radiation tolerance of modern CMOS processes and hardware/software mitigation techniques. Rather than focusing on specific device-level hardening parameters, we emphasize architectural strategies that enable robust operation using radiation-tolerant commercial technologies.

The combination of inherent CMOS radiation tolerance improvements and layered mitigation strategies makes co-flying standard CMOS compute viable for HWO. We recommend a defense-in-depth approach:

\paragraph{Component Selection}
Prefer processors fabricated in FinFET or \gls{gaa} processes on \gls{soi} substrates where available. Modern data
center GPUs already incorporate many of these features for reliability reasons. 

\paragraph{Memory Protection} Employ \gls{ecc} memory throughout. Since HOWFSC accesses all matrix data each iteration, normal memory access will trigger error correction without dedicated scrubbing.

\paragraph{Computational Integrity} Implement \gls{abft} checksums for EFC matrix operations, with automatic recomputation on detected errors.

\paragraph{Control Path Redundancy} Apply \gls{tmr} to critical control components that cannot be protected by \gls{abft}, such as state machines and communication interfaces. 
        
\paragraph{Graceful Degradation}Design the control system to continue operation with reduced frequency if radiation events cause temporary compute unavailability.

This layered approach allows the co-flying satellite to leverage the performance advantages of standard CMOS hardware while achieving reliability appropriate for supporting HWO's science mission.

\section{Conclusion}
In this work, we show that the central obstacle to achieving high-frequency HOWFSC in space for the HWO is its extreme memory bandwidth and capacity requirements at LUVOIR-scale. Radiation-hardened processors fall short by several orders of magnitude, and GITL operation cannot reach the necessary control rates due to inherent latency.

We conduct a system-level exploration of compute strategies to address this challenge. Using roofline-based analysis, we find that only bleeding edge compute systems can approach the required performance, and only when computation is relocated from Earth to co-flying spacecraft capable of supporting radiation-tolerant and power-efficient platforms.

We also present initial concepts of an HBM-based process-near-memory design and a distributed SRAM architecture that can support HOWFSC cadences up to 300 Hz while reducing power consumption significantly, which in turn enables lowered computational satellite design, launch, and operating costs. These results establish co-flying compute as a practical and enabling approach for continuous dark hole maintenance on future flagship missions.

By connecting the computational requirements of advanced coronagraphy with emerging compute system architectures, this work provides a foundation for overcoming the current memory and communications bottlenecks, enabling practical high-frequency wavefront control for HWO.
Future work should refine architectural designs beyond high-level modeling, including physical layout, packaging, and interconnect constraints. Additional research is also needed to evaluate alternative numerical formulations, reduced-precision variants of HOWFSC algorithms, and the integration of layered fault tolerance under realistic radiation environments. These efforts will be essential for determining the optimal balance of performance, energy, reliability, and mission cost.

\appendix
\section{Coronagraph System Parameters}
\label{sec:CPARAM}

Table \ref{tab:coronagraph_param} shows the coronagraph parameters used for our analysis, adapted from Pogorelyuk et al \cite{COMPLEX}. We assume a 128$\times$128 DM array and a precomputed matrix stored in double precision to best approximate worst-case requirements.

Table \ref{tab:abbrev} shows some frequently used abbreviations in deriving roofline models.

\begin{table}[t]
\centering
\begin{tabular}{ccccc}
\hline
\cellcolor{gray!20} Coronagraph & \cellcolor{gray!20} \# Channel & \cellcolor{gray!20} IWA(L/D) & \cellcolor{gray!20} OWA (L/D) & \cellcolor{gray!20}\# DM Actuator \\
\hline
LUVOIR VIS A
  & 5
  & 3.5
  & 64
  & 128$\times$128 \\
\hline
\cellcolor{gray!20}\# Pixel/Side & \cellcolor{gray!20} L/D per Pixel & \cellcolor{gray!20} Dark Hole Area & \cellcolor{gray!20} \# Active Pixel & \cellcolor{gray!20}\# Active Actuator \\
\hline
  \rule[-1ex]{0pt}{3.5ex} 400
  & 0.5
  & \SI{12829}{\milli\meter\squared}
  & 51316
  & 25736 \\
\hline
  \cellcolor{gray!20}\# Probe & \cellcolor{gray!20} Pixel Bit-Depth & \cellcolor{gray!20} \# \cellcolor{gray!20} Gain Size & \cellcolor{gray!20} Jacobian Size & \cellcolor{gray!20}\# E Field States \\
\hline
  \rule[-1ex]{0pt}{3.5ex} 7 
  & 12 
  & \SI{5.30}{\giga\byte}
  & \SI{105.65}{\giga\byte} 
  & 513160\\
\hline
\end{tabular}
\vspace{0.5em}
\caption{Coronagraph Parameters Used\cite{COMPLEX}}
\label{tab:coronagraph_param}
\vspace{1em}
\end{table}

\begin{table}[t]
\centering
\begin{tabular}{cccccc}
\hline
\rowcolor{gray!20}\# Channel & \cellcolor{gray!20} \# Pixels/Side & \cellcolor{gray!20} \# Active Pixels & \cellcolor{gray!20} \# Active Actuator & \cellcolor{gray!20} \# Probe & \cellcolor{gray!20}  E Field States\\
\hline
  \rule[-1ex]{0pt}{3.5ex} $N_{ch}$
  & $N_{side}$
  & $N_{pix}$
  & $N_{act}$
  & $N_{Probe}$ 
  & $M$\\
\hline
\end{tabular}
\vspace{0.5em}
\caption{Math Notations and Abbreviations}
\label{tab:abbrev}
\end{table}

We derive active pixels and actuators using the following formulas (Section 7 of \citen{COMPLEX}). 
\begin{align}
     N_{act} &\approx 2\cdot\pi\left( \frac{N_{side}}{2} \right)^2\\
    N_{pix} &\approx \frac{4\pi}{\lambda/D}\cdot(\text{OWA}^2-\text{IWA}^2)
\end{align}

We derive the Jacobian size and image size using the following formulas \cite{COMPLEX}.
\begin{align}
    \text{Jacobian Size} &= 2N_{act} N_{pix} N_{ch} \cdot \text{Bytes/Float} = 105.65~\text{GB}\\
    \text{Image Size} &= N_{pix} N_{ch} \cdot \text{Pixel Bit-Depth}/8 = 375.85~\text{KB}
\end{align}

\section{Hardware Design Parameters}
\subsection{Interconnection Network}
 \label{sec:itcbw}
Considering that the size of the input electric field $\mathbf{E}$ is $16N_{pix}N_{channel}$ bytes and both the size of the intermediate product $\mathbf J^T\mathbf E$ and the state vector $\Delta\mathbf u$ are $8 N_{act}$ bytes, Eq. \ref{eq:BROAD} and \ref{eq:GATHER} show the broadcast and gather latencies for $N$ processors connected via a $d$-degree H-Tree network with point-to-point bandwidths of $BW$. 
\begin{align}
    \label{eq:BROAD}
    \text{Broadcast Latency} &=  \log_d N \frac{16N_{pix}N_{channel}+8N_{act}}{BW}\\
    \label{eq:GATHER}
    \text{Gather Latency} &= \sum_{k=1}^{\log_d N} \frac{16N_{act}}{d^k BW} \leq \frac{16N_{act}}{BW}
\end{align}

We analyze the feasibility of such a communication network by computing the required point-to-point bandwidth for HOWFSC at control frequency $f$ (Eq. \ref{eq:PBW}), assuming that 10\% of the control cadence is allocated to communication:
\begin{align}
    \label{eq:PBW}
    \text{Point-to-Point BW} &= 10f\left((16N_{pix}N_{channel}+8N_{act})\log_d{N} + 16N_{act}\right)
\end{align}

\section{Radiation Tolerance}
\label{sec:rad_appx}
\subsection{Inherent Radiation Tolerance of Modern CMOS}

Modern standard CMOS \emph{logic} technologies have become increasingly radiation-tolerant as a byproduct of device scaling, even without explicit hardening. This trend primarily benefits logic transistors used in CPUs, GPUs, and ASICs. In contrast, DRAM and HBM remain more radiation-sensitive because stored charge scales poorly and is more easily upset by ionizing particles.

\paragraph{\Gls{soi}}
\Gls{soi} substrates isolate the active device region with a buried oxide, reducing charge collection and eliminating latchup paths, which significantly improves \gls{seu} and \gls{sel} tolerance\cite{SOI_SEL, SOI_RAD}.

\paragraph{FinFET Architecture}
FinFET devices reduce sensitive volume and provide stronger electrostatic control than planar CMOS, yielding roughly an order-of-magnitude reduction in \gls{seu} rate at \SI{16}{\nano\meter} and improved \gls{tid} robustness\cite{FINFET_RAD, FINFET_SEU}.

\paragraph{\Gls{gaa} Architecture}
\Gls{gaa} nanosheet transistors further shrink charge-collection volume and improve threshold stability under irradiation. Early results show superior \gls{tid} response compared to FinFETs, and \SI{2}{\nano\meter}-class \gls{gaa} nodes expected in the 2040s should provide sufficient inherent tolerance for many space applications\cite{GAA_RAD, GAA_TID}.

\subsubsection{Increasing Adoption of Standard CMOS in Space}
These improvements have accelerated the adoption of standard CMOS in space systems, including CubeSats and LEO constellations, where processors, GPUs, and memory devices fabricated in commercial processes operate reliably with appropriate system-level mitigation\cite{COTS_NEWSPACE, COTS_SPACE}.

\subsection{Hardware Mitigation Strategies}

Even with improved inherent tolerance, standard CMOS hardware requires additional protection for high-reliability missions. The following hardware-level strategies form the foundation for reliable operation.

\paragraph{Physical Shielding}
Aluminum shielding of \SIrange{2}{5}{\milli\meter} can attenuate low-energy protons and electrons, and spot shielding can protect vulnerable components such as voltage regulators and configuration memory. However, shielding has diminishing returns against high-energy \glspl{gcr} and increases spacecraft mass\cite{SOI_RAD}.

\paragraph{Redundancy Architectures}
\Gls{tmr} remains the standard approach for mitigating \glspl{seu} in critical logic. Three parallel modules vote on outputs to mask faults\cite{TMR_SPACE}. In HOWFSC, \gls{tmr} is most effective for control paths and state machines, while matrix operations are better protected through algorithm-based fault tolerance (Section \ref{subsubsec:abft}).

\paragraph{Error-Correcting Memory}
\Gls{ecc} memory is essential for protecting storage and data paths. DRAM and HBM typically support single-error-correction, double-error-detection(SECDED) or chipkill, while SRAM-based designs allow stronger custom \gls{ecc} tailored to expected error rates. Radiation testing of Google’s Trillium TPU showed HBM to be the most sensitive subsystem, exhibiting irregularities after only \SI{2}{\kilo\rad(Si)}\cite{SUNCATCHER}, underscoring the need for robust \gls{ecc} and highlighting the benefits of SRAM-based custom architectures.

\subsection{Software and System-Level Mitigation}

Software techniques complement hardware protection by providing recovery
mechanisms and algorithmic resilience.

\subsubsection{\Gls{abft}}
\label{subsubsec:abft}
\Gls{abft} embeds error detection and correction directly into numerical algorithms by maintaining checksum relationships across matrix elements\cite{ABFT}. This makes it highly effective for linear operations and therefore well suited to HOWFSC workloads. For the GEMV operations that dominate EFC, row and column checksums allow errors to be detected and located at the end of each computation, enabling immediate recomputation when necessary. The overhead of \gls{abft} scales only with matrix dimensions rather than the total number of elements, resulting in minimal cost even for large matrices, and the technique provides protection against both memory and computational faults. Unlike generic redundancy schemes, \gls{abft} leverages the mathematical structure of linear algebra to achieve efficient fault coverage, and it can be scaled by increasing redundant computation during high-radiation periods such as solar flares without requiring hardware changes. Because HOWFSC naturally reads all matrix data every iteration, dedicated memory scrubbing is unnecessary—\gls{ecc} will detect and correct single-event upsets during normal access.

\subsubsection{Checkpointing and Rollback}
Periodic checkpointing of system state enables recovery from detected errors by
rolling back to a known-good state\cite{CHECKPOINT}. For HOWFSC, natural
checkpoint boundaries exist between control iterations: the DM commands from
each successful iteration can be checkpointed, allowing the system to recover
from a corrupted computation by reverting to the previous valid state and
recomputing.

\section{Model Validation}
\label{sec:validation}
We validate our roofline model using a cloud instance equipped with an NVIDIA B200 GPU. The core linear-algebra operations of electric field conjugation (EFC) are implemented in PyTorch and executed using the CUDA backend\cite{PYTORCH}. Measured execution times are obtained through runtime profiling and compared against the model predictions. 

For LUVOIR-A EFC, the model predicts a runtime of \SI{14.4}{\milli\second}, while the measured execution time is \SI{17.7}{\milli\second}, corresponding to an error of approximately 23\%. We attribute this discrepancy primarily to kernel launch overheads and Python runtime overheads in the PyTorch implementation. A more optimized implementation using fused or custom CUDA kernels would be expected to reduce this gap and more closely approach the modeled performance.

We also measure GPU power consumption using the \texttt{nvidia-smi} monitoring tool. The GPU exhibits an idle power of approximately \SI{150}{\watt}. During EFC execution, power consumption remains near the device thermal design power (TDP) of \SI{1000}{\watt} for the duration of the computation. As a result, the energy consumed per EFC iteration is dominated by static and memory-system power rather than useful arithmetic work. This behavior highlights a fundamental inefficiency of large, throughput-optimized GPUs for EFC workloads, which are memory-bound and latency-sensitive, and motivates the exploration of more energy-efficient, memory-centric hardware architectures for high-frequency HOWFSC.

\subsection*{Disclosures}

The authors declare no conflicts of interest.

\subsection* {Code and Data Availability} 
All code necessary to reproduce the results in this paper are public available at the repository\\\href{https://github.com/fangzhonglyu/HOWFSC}{https://github.com/fangzhonglyu/HOWFSC}. Hardware synthesis experiments depend on proprietary EDA tools and therefore require separate commercial licenses.

\subsection*{Acknowledgments}
The authors are especially thankful to Jamila Taaki, Leonid Pogorelyuk, Laurent Pueyo, and Kerri Cahoy for helpful discussions regarding wavefront sensing and control algorithms throughout this research. ChatGPT and Claude were used for writing refinement and grammar checking.

\bibliography{report}   

\begin{thebibliography}{10}

\bibitem{EFF_DRIFT}
Pogorelyuk, L., Pueyo, L., and Kasdin, N.~J., ``{On the effects of pointing jitter, actuator drift, telescope rolls, and broadband detectors in dark hole maintenance and electric field order reduction},'' {\em Journal of Astronomical Telescopes, Instruments, and Systems}~{\bf 6}(3),  039001 (2020).

\bibitem{COMPLEX}
Pogorelyuk, L., Haughwout, C., Belsten, N., Cady, E., and Cahoy, K., ``{Computational complexities of image plane algorithms for high contrast imaging in space telescopes},'' {\em Journal of Astronomical Telescopes, Instruments, and Systems}~{\bf 8}(4),  049003 (2022).

\bibitem{EMBEDDED}
Belsten, N., Milani, K., Pogorelyuk, L., Eickert, B., Rao, S., Douglas, E.~S., and Cahoy, K., ``{Evaluating embedded hardware for high-order wavefront sensing and control},'' in [{\em Techniques and Instrumentation for Detection of Exoplanets XI}{\nolinebreak\hspace{0.1em}]},  Ruane, G.~J., ed.,  {\bf 12680},  126801N, International Society for Optics and Photonics, SPIE (2023).

\bibitem{THESIS}
Belsten, N., {\em Embedded Computing for Wavefront Control on Future Space Telescopes}, PhD thesis, Massachusetts Institute of Technology (2025).

\bibitem{DZM_OG}
Pogorelyuk, L. and Kasdin, N.~J., ``Dark hole maintenance and a posteriori intensity estimation in the presence of speckle drift in a high-contrast space coronagraph,'' {\em The Astrophysical Journal}~{\bf 873},  95 (Mar. 2019).

\bibitem{DZM_IMPL}
Redmond, S.~F., Pogorelyuk, L., Pueyo, L., Por, E., Noss, J., Will, S.~D., Laginja, I., Brooks, K., Maclay, M., Fowler, J., Kasdin, N.~J., Perrin, M.~D., and Soummer, R., ``{Implementation of a dark zone maintenance algorithm for speckle drift correction in a high contrast space coronagraph},'' {\em Journal of Astronomical Telescopes, Instruments, and Systems}~{\bf 8}(3),  035001 (2022).

\bibitem{EFC_CITE}
Give'on, A., Kern, B., Shaklan, S., Moody, D.~U., and Pueyo, L., ``{Broadband wavefront correction algorithm for high-contrast imaging systems},'' in [{\em Astronomical Adaptive Optics Systems and Applications III}{\nolinebreak\hspace{0.1em}]},  Tyson, R.~K. and Lloyd-Hart, M., eds.,  {\bf 6691},  66910A, International Society for Optics and Photonics, SPIE (2007).

\bibitem{LDFC}
Miller, K., Guyon, O., and Males, J., ``{Spatial linear dark field control: stabilizing deep contrast for exoplanet imaging using bright speckles},'' {\em Journal of Astronomical Telescopes, Instruments, and Systems}~{\bf 3}(4),  049002 (2017).

\bibitem{ADJOINT}
Milani, K., Will, S.~D., Gorkom, K. J.~V., Douglas, E.~S., Ashcraft, J.~N., and Cahoy, K.~L., ``{Demonstrations of adjoint electric field conjugation for a vortex coronagraph},'' {\em Journal of Astronomical Telescopes, Instruments, and Systems}~{\bf 11}(3),  039001 (2025).

\bibitem{PWP}
Allan, G., Ruane, G., Walter, A.~B., Riggs, A. J.~E., Prada, C.~M., Noyes, M., Poon, P.~K., Llop-Sayson, J.~D., Jovanovic, N., Coker, C., Stark, C., Mennesson, B., and Mawet, D., ``{Demonstration of coronagraph technology for high-contrast point spectroscopy of ExoEarths},'' in [{\em Techniques and Instrumentation for Detection of Exoplanets XI}{\nolinebreak\hspace{0.1em}]},  Ruane, G.~J., ed.,  {\bf 12680},  1268019, International Society for Optics and Photonics, SPIE (2023).

\bibitem{MODAL}
Manojkumar, S., Redmond, S.~F., Pogorelyuk, L., Page, C.~L., Gill, A.~S., Pueyo, L., Por, E.~H., Laginja, I., Pourcelot, R., Nickson, B.~F., Sahoo, A., Nguyen, M.~M., Soummer, R., Perrin, M.~D., Nemati, B., Kasdin, J.~N., and Cahoy, K., ``{Modal estimation of high-order wavefront drifts on coronagraphic testbeds for high-contrast direct imaging of exoplanets},'' in [{\em Techniques and Instrumentation for Detection of Exoplanets XII}{\nolinebreak\hspace{0.1em}]},  Ruane, G.~J. and Millar-Blanchaer, M.~A., eds.,  {\bf 13627},  1362716, International Society for Optics and Photonics, SPIE (2025).

\bibitem{OPTICAL_PROP}
Sun, H., Kasdin, N.~J., and Vanderbei, R., ``{Identification and adaptive control of a high-contrast focal plane wavefront correction system},'' {\em Journal of Astronomical Telescopes, Instruments, and Systems}~{\bf 4}(4),  049006 (2018).

\bibitem{ROMAN_DEMO}
Mennesson, B., Bailey, V.~P., Zellem, R., Hildebrandt, S., Ygouf, M., Rhodes, J., Zimmerman, N., Nemati, B., Gonzalez, G., Cady, E., Kern, B., Koch, T., Krist, J., Heydorff, K., Luchik, T., Mok, F., Morrissey, P., Poberezhskiy, I., Riggs, A.~J., Shi, F., Zhao, F., Akeson, R., Armus, L., Greenbaum, A., Ingalls, J., and Lowrance, P., ``{The Roman Space Telescope coronagraph technology demonstration: current status and relevance to future missions},'' in [{\em Space Telescopes and Instrumentation 2022: Optical, Infrared, and Millimeter Wave}{\nolinebreak\hspace{0.1em}]},  Coyle, L.~E., Matsuura, S., and Perrin, M.~D., eds.,  {\bf 12180},  121801W, International Society for Optics and Photonics, SPIE (2022).

\bibitem{PUEYO_PRES}
Pueyo, L.~A., Pogorelyuk, L., Laginja, I., Soummer, R., Sahoo, A., Por, E., Cahoy, K., Coyle, L., and Knight, S., ``{Architecture trades to optimize wavefront stability requirements for exoplanet imaging in space.},'' in [{\em Space Telescopes and Instrumentation 2022: Optical, Infrared, and Millimeter Wave}{\nolinebreak\hspace{0.1em}]},  Coyle, L.~E., Matsuura, S., and Perrin, M.~D., eds.,  {\bf 12180},  121802J, International Society for Optics and Photonics, SPIE (2022).

\bibitem{CUBESAT}
Schieler, C.~M., Riesing, K.~M., Horvath, A.~J., Bilyeu, B.~C., Chang, J.~S., Garg, A.~S., Wang, J.~P., and Robinson, B.~S., ``{200 Gbps TBIRD CubeSat downlink: pre-flight test results},'' in [{\em Free-Space Laser Communications XXXIV}{\nolinebreak\hspace{0.1em}]},  Hemmati, H. and Robinson, B.~S., eds.,  {\bf 11993},  119930P, International Society for Optics and Photonics, SPIE (2022).

\bibitem{starcloud}
Feilden, E., Oltean, A., and Johnston, P., ``Why we should train ai in space,'' white paper, Starcloud, Inc. (2024).
\newblock White Paper v1.03, September 2024.

\bibitem{SUNCATCHER}
y~Arcas, B.~A., Beals, T., Biggs, M., Bloom, J.~V., Fischbacher, T., Gromov, K., K\"{o}ster, U., Pravahan, R., and Manyika, J., ``Towards a future space-based, highly scalable {AI} infrastructure system design,'' tech. rep., Google (2025).
\newblock Project Suncatcher Whitepaper.

\bibitem{STARLINK_DISR}
Foust, J., ``The space review: Starlink’s disruption of the space industry,'' (May 2024).

\bibitem{DSN}
{NASA Deep Space Network}, ``Dsn services catalog 820-100, rev. h,'' technical report, Jet Propulsion Laboratory / NASA (2022).
\newblock DSN Services Catalog; Deep Space Network.

\bibitem{MEMS_PERF}
Zhang, C.~C., Foster, W.~B., Downey, R.~D., Arrasmith, C.~L., and Dickensheets, D.~L., ``{Dynamic performance of MEMS deformable mirrors for use in an active/adaptive two-photon microscope},'' in [{\em Adaptive Optics and Wavefront Control for Biological Systems II}{\nolinebreak\hspace{0.1em}]},  Bifano, T.~G., Kubby, J., and Gigan, S., eds.,  {\bf 9717},  97170G, International Society for Optics and Photonics, SPIE (2016).

\bibitem{MEMS_HCI}
Morgan, R.~E., Douglas, E.~S., Allan, G.~W., Bierden, P., Chakrabarti, S., Cook, T., Egan, M., Furesz, G., Gubner, J.~N., Groff, T.~D., Haughwout, C.~A., Holden, B.~G., Mendillo, C.~B., Ouellet, M., do~Vale~Pereira, P., Stein, A.~J., Thibault, S., Wu, X., Xin, Y., and Cahoy, K.~L., ``Mems deformable mirrors for space-based high-contrast imaging,'' {\em Micromachines}~{\bf 10}(6) (2019).

\bibitem{ROOFLINE}
Williams, S., Waterman, A., and Patterson, D., ``Roofline: an insightful visual performance model for multicore architectures,'' ~{\bf 52}(4) (2009).

\bibitem{BAE5545}
{BAE Systems}, {\em RAD5545 SpaceVPX SBC v1.2 Flight – Data Sheet}.
\newblock BAE Systems (Jan. 2025).
\newblock SpaceVPX single-board computer for space applications.

\bibitem{LX2160A}
{Teledyne e2v Semiconductors SAS}, {\em LX2160-Space: Radiation-Tolerant 16× ARM Cortex-A72 2.2 GHz Microprocessor \\ Datasheet – Preliminary Specification}.
\newblock Teledyne e2v (Mar. 2023).
\newblock Preliminary specification.

\bibitem{orinnx}
{Nvidia Corporation}, {\em Jetson Orin NX Series Modules Data Sheet (DS-10712-001, v1.6)}.
\newblock Nvidia (2025).
\newblock Jetson Orin NX system-on-module specifications.

\bibitem{H100}
{Nvidia Corporation}, {\em NVIDIA H100 Tensor Core GPU Data Sheet}.
\newblock Nvidia (2024).
\newblock H100 Tensor Core GPU specifications.

\bibitem{9575F}
{AMD}, ``Amd epyc™ 9575f processor.'' \url{https://www.amd.com/en/products/processors/server/epyc/9005-series/amd-epyc-9575f.html} (2025).
\newblock Accessed: 2025-11-28.

\bibitem{AETHERO}
{Aethero}, ``Aethero — space data, re-imagined.'' \url{https://www.aethero.com/} (2025).
\newblock Accessed: 2025-11-28.

\bibitem{EPYC_PERF}
{AMD}, ``Leadership hpc performance with 5th generation amd epyc processors.'' \url{https://www.amd.com/en/blogs/2025/leadership-hpc-performance-with-5th-generation-amd.html} (Jan. 2025).
\newblock Accessed: 2025-11-28.

\bibitem{B200}
{NVIDIA Corporation}, {\em NVIDIA Blackwell Architecture Data Sheet}.
\newblock NVIDIA (2025).

\bibitem{MI300X}
{AMD}, {\em AMD Instinct MI300X Accelerator Data Sheet}.
\newblock AMD (2025).
\newblock Discrete GPU accelerator for AI and HPC applications.

\bibitem{TPUv4}
Jouppi, N., Kurian, G., Li, S., Ma, P., Nagarajan, R., Nai, L., Patil, N., Subramanian, S., Swing, A., Towles, B., Young, C., Zhou, X., Zhou, Z., and Patterson, D.~A., ``Tpu v4: An optically reconfigurable supercomputer for machine learning with hardware support for embeddings,'' {\em ISCA '23}, Association for Computing Machinery, New York, NY, USA (2023).

\bibitem{B300}
{NVIDIA Corporation}, {\em NVIDIA Blackwell Ultra GPU Architecture Data Sheet}.
\newblock NVIDIA (2025).
\newblock Blackwell Ultra architecture specifications.

\bibitem{SPACE_DC}
Bleier, N., Mubarik, M.~H., Swenson, G.~R., and Kumar, R., ``Space microdatacenters,'' {\em MICRO '23},  900–915, Association for Computing Machinery, New York, NY, USA (2023).

\bibitem{2NM_SRAM}
Chang, T.-Y.~J., Chen, Y.-H., Reddy, K.~V., Puri, N., Masina, T., Lin, K.-C., Wang, P.-S., Lin, Y., Lin, C.-Y., Nien, Y.-H., Fujiwara, H., Lin, K.-F., Chang, M.-H., Wu, C.~W., Lee, R., Wang, Y., Liao, H.-J., Li, Q., Wang, P.~W., and Yeap, G., ``A 38.1mb/mm2 sram in a 2nm-cmos-nanosheet technology for high-density and energy-efficient compute,'' in [{\em 2025 IEEE International Solid-State Circuits Conference (ISSCC)}{\nolinebreak\hspace{0.1em}]},   {\bf 68},  492--494 (2025).

\bibitem{RRAM}
Jiang, Z., Qin, S., Li, H., Fujii, S., Lee, D., Wong, S., and Wong, H.-S.~P., ``Next-generation ultrahigh-density 3-d vertical resistive switching memory (vrsm)—part ii: Design guidelines for device, array, and architecture,'' {\em IEEE Transactions on Electron Devices}~{\bf 66}(12),  5147--5154 (2019).

\bibitem{AMBER_RRAM}
Upton, L.~R., Levy, A., Scott, M.~D., Rich, D., Khwa, W.-S., Chih, Y.-D., Chang, M.-F., Mitra, S., Raina, P., and Murmann, B., ``Ember: A 100 mhz, 0.86 mm2, multiple-bits-per-cell rram macro in 40 nm cmos with compact peripherals and 1.0 pj/bit read circuitry,'' in [{\em ESSCIRC 2023- IEEE 49th European Solid State Circuits Conference (ESSCIRC)}{\nolinebreak\hspace{0.1em}]},   469--472 (2023).

\bibitem{MRAM}
Chang, T.-C., Chiu, Y.-C., Lee, C.-Y., Hung, J.-M., Chang, K.-T., Xue, C.-X., Wu, S.-Y., Kao, H.-Y., Chen, P., Huang, H.-Y., Teng, S.-H., and Chang, M.-F., ``13.4 a 22nm 1mb 1024b-read and near-memory-computing dual-mode stt-mram macro with 42.6gb/s read bandwidth for security-aware mobile devices,'' in [{\em 2020 IEEE International Solid-State Circuits Conference - (ISSCC)}{\nolinebreak\hspace{0.1em}]},   224--226 (2020).

\bibitem{HBM}
Chae, K., Song, J., Choi, Y., Park, J., Koo, B., Oh, J., Yi, S., Lee, W., Kim, D., Kang, K., Kim, E., Kim, J., Park, S., Park, S., Noh, M., Rhew, H.~G., and Shin, J., ``A 4-nm 1.15 tb/s hbm3 interface with resistor-tuned offset calibration and in situ margin detection,'' {\em IEEE Journal of Solid-State Circuits}~{\bf 59}(1),  231--242 (2024).

\bibitem{HBM_PHY}
Choi, J., Kim, Y.-G., Kim, J., Chung, J., Jeon, Y.-D., Cho, M.-H., Park, S., and Han, J., ``A 6.4gb/s/pin hbm3 digital phy with low-power, area-efficient techniques for chiplet-based ai processors in 12-nm cmos,'' in [{\em 2024 IEEE Asian Solid-State Circuits Conference (A-SSCC)}{\nolinebreak\hspace{0.1em}]},   1--3 (2024).

\bibitem{DDR}
Kim, C., Lee, H.-W., and Song, J., ``Memory interfaces: Past, present, and future,'' {\em IEEE Solid-State Circuits Magazine}~{\bf 8}(2),  23--34 (2016).

\bibitem{NAND}
Jeong, W., Im, J.-w., Kim, D.-H., Nam, S.-W., Shim, D.-K., Choi, M.-H., Yoon, H.-J., Kim, D.-H., Kim, Y.-S., Park, H.-W., Kwak, D.-H., Park, S.-W., Yoon, S.-M., Hahn, W.-G., Ryu, J.-H., Shim, S.-W., Kang, K.-T., Ihm, J.-D., Kim, I.-M., Lee, D.-S., Cho, J.-H., Kim, M.-S., Jang, J.-H., Hwang, S.-W., Byeon, D.-S., Yang, H.-J., Park, K., Kyung, K.-H., and Choi, J.-H., ``A 128 gb 3b/cell v-nand flash memory with 1 gb/s i/o rate,'' {\em IEEE Journal of Solid-State Circuits}~{\bf 51}(1),  204--212 (2016).

\bibitem{INTERCON}
De~Sensi, D., Pichetti, L., Vella, F., De~Matteis, T., Ren, Z., Fusco, L., Turisini, M., Cesarini, D., Lust, K., Trivedi, A., Roweth, D., Spiga, F., Di~Girolamo, S., and Hoefler, T., ``Exploring gpu-to-gpu communication: Insights into supercomputer interconnects,'' in [{\em Proceedings of the International Conference for High Performance Computing, Networking, Storage, and Analysis}{\nolinebreak\hspace{0.1em}]},  {\em SC '24}, IEEE Press (2024).

\bibitem{STARLINK_MINILSR}
{SpaceX}, ``Starlink mini laser terminals.'' \url{https://www.starlink.com/technology}.
\newblock Accessed: 29 Nov 2025.

\bibitem{bleier2025architecting}
Bleier, N., Eason, R., Lembeck, M., and Kumar, R., ``Architecting space microdatacenters: A system-level approach,'' {\em 2025 IEEE International Symposium on High Performance Computer Architecture (HPCA)}  (2025).

\bibitem{SRAM_LEAK}
Wang, J.-S., Liu, C.-T., and Hou, Y.-C., ``A 0.5-v 125-mhz 256-kb 22-nm sram with 10-aj/bit active energy and 10-pw/bit shutdown power,'' {\em IEEE Journal of Solid-State Circuits}~{\bf 60}(8),  3043--3052 (2025).

\bibitem{ASAP7}
Clark, L.~T., Vashishtha, V., Shifren, L., Gujja, A., Sinha, S., Cline, B., Ramamurthy, C., and Yeric, G., ``Asap7: A 7-nm finfet predictive process design kit,'' {\em Microelectronics Journal}~{\bf 53},  105--115 (2016).

\bibitem{PARALLELIO}
Mansuri, M., Jaussi, J.~E., Kennedy, J.~T., Hsueh, T.-C., Shekhar, S., Balamurugan, G., O'Mahony, F., Roberts, C., Mooney, R., and Casper, B., ``A scalable 0.128–1 tb/s, 0.8–2.6 pj/bit, 64-lane parallel i/o in 32-nm cmos,'' {\em IEEE Journal of Solid-State Circuits}~{\bf 48}(12),  3229--3242 (2013).

\bibitem{UCIE_COST}
Das~Sharma, D., Pasdast, G., Qian, Z., and Aygun, K., ``Universal chiplet interconnect express (ucie): An open industry standard for innovations with chiplets at package level,'' {\em IEEE Transactions on Components, Packaging and Manufacturing Technology}~{\bf 12}(9),  1423--1431 (2022).

\bibitem{HW_NEURON}
Liu, Y., Chen, Y., Zhao, Y., Hao, Y., Zheng, Z., Kong, W., Li, Z., Jiang, D., Xia, R., Ma, Z., Liu, Z., Wan, Z., Lu, Y., Liu, X., Guo, H., Yang, Z., Wang, Z., Ma, T., Zou, M., Zhang, R., Li, L., Hu, X., Du, Z., Xu, Z., Guo, Q., Chen, T., and Chen, Y., ``Hardwired-neurons language processing units as general-purpose cognitive substrates,'' (2025).

\bibitem{JWST_RAD}
Greenholt, T.~J., Figer, D.~F., Rauscher, B.~J., Regan, M., Moseley, S.~H., and Garnett, J., ``Radiation environment performance of {JWST} prototype {FPAs},'' in [{\em Space Telescopes and Instrumentation 2008: Optical, Infrared, and Millimeter}{\nolinebreak\hspace{0.1em}]},   {\bf 7010},  70103U (2008).

\bibitem{L2_RAD}
Mewaldt, R.~A., Davis, A.~J., Lave, K.~A., Leske, R.~A., Stone, E.~C., Wiedenbeck, M.~E., Binns, W.~R., Christian, E.~R., Cummings, A.~C., de~Nolfo, G.~A., Israel, M.~H., Labrador, A.~W., and von Rosenvinge, T.~T., ``Record-setting cosmic-ray intensities in 2009 and 2010,'' {\em The Astrophysical Journal Letters}~{\bf 723}(1),  L1 (2010).

\bibitem{NGST_RAD}
Barth, J.~L., Isaacs, J.~C., and Poivey, C., ``The radiation environment for the next generation space telescope,'' in [{\em Proceedings of the RADECS Conference}{\nolinebreak\hspace{0.1em}]},  (1999).

\bibitem{SOI_SEL}
Schwank, J.~R., Ferlet-Cavrois, V., Shaneyfelt, M.~R., Paillet, P., and Dodd, P.~E., ``Radiation effects in {SOI} technologies,'' {\em IEEE Transactions on Nuclear Science}~{\bf 50}(3),  522--538 (2003).

\bibitem{SOI_RAD}
Schwank, J.~R., Shaneyfelt, M.~R., and Dodd, P.~E., ``Radiation hardness assurance testing of microelectronic devices and integrated circuits: Radiation environments, physical mechanisms, and foundations for hardness assurance,'' {\em IEEE Transactions on Nuclear Science}~{\bf 60}(3),  2074--2100 (2013).

\bibitem{FINFET_RAD}
King, M.~P., Schrimpf, R.~D., Massengill, L.~W., Fleetwood, D.~M., Pellish, J.~A., Reed, R.~A., Weller, R.~A., Mendenhall, M.~H., and Edmonds, L.~D., ``{FinFET} technologies for digital systems with radiation requirements: {TID}, {SEE}, basic mechanisms and lessons learnt,'' in [{\em CERN FinFET Workshop}{\nolinebreak\hspace{0.1em}]},  (2017).

\bibitem{FINFET_SEU}
Seifert, N., Gill, B., Jahinuzzaman, S., Basile, J., Ambrose, V., Shi, Q., Allmon, R., and Bramnik, A., ``Soft error susceptibilities of 22 nm tri-gate devices,'' {\em IEEE Transactions on Nuclear Science}~{\bf 59}(6),  2666--2673 (2012).

\bibitem{GAA_RAD}
Kumar, P., Koley, K., and Kumar, S., ``Radiation effects on advanced multi-gate 3d devices and materials,'' {\em Transactions on Electrical and Electronic Materials}  (2025).

\bibitem{GAA_TID}
Colinge, J.~P., Quinn, A.~J., Floyd, L., Duerinckx, G., and Groeseneken, G., ``Effects of total-dose irradiation on gate-all-around {(GAA)} devices,'' {\em IEEE Transactions on Nuclear Science}~{\bf 39}(5),  1082--1087 (1992).

\bibitem{COTS_NEWSPACE}
Poizat, M., Pangaud, A., and Beutier, T., ``Toward the use of electronic commercial off-the-shelf devices in space: Assessment of the true radiation environment in low earth orbit {(LEO)},'' {\em Electronics}~{\bf 12}(19),  4058 (2023).

\bibitem{COTS_SPACE}
LaBel, K.~A., Gates, M.~M., Moran, A.~K., Marshall, P.~W., Barth, J., Stassinopoulos, E.~G., Seidleck, C.~M., and Dale, C.~J., ``Commercial microelectronics technologies for applications in the satellite radiation environment,''  375--390 (1996).

\bibitem{TMR_SPACE}
Pratt, B., Caffrey, M., Carroll, J.~F., Graham, P., Morgan, K., and Wirthlin, M., ``Fine-grain {SEU} mitigation for {FPGAs} using partial {TMR},'' {\em IEEE Transactions on Nuclear Science}~{\bf 55}(4),  2274--2280 (2008).

\bibitem{ABFT}
Huang, K.~H. and Abraham, J.~A., ``Algorithm-based fault tolerance for matrix operations,'' {\em IEEE Transactions on Computers}~{\bf C-33}(6),  518--528 (1984).

\bibitem{CHECKPOINT}
Elnozahy, E.~N., Alvisi, L., Wang, Y.-M., and Johnson, D.~B., ``A survey of rollback-recovery protocols in message-passing systems,'' {\em ACM Computing Surveys}~{\bf 34}(3),  375--408 (2002).

\bibitem{PYTORCH}
Paszke, A., Gross, S., Massa, F., Lerer, A., Bradbury, J., Chanan, G., Killeen, T., Lin, Z., Gimelshein, N., Antiga, L., Desmaison, A., K{\"{o}}pf, A., Yang, E.~Z., DeVito, Z., Raison, M., Tejani, A., Chilamkurthy, S., Steiner, B., Fang, L., Bai, J., and Chintala, S., ``Pytorch: An imperative style, high-performance deep learning library,'' {\em CoRR}~{\bf abs/1912.01703} (2019).

\end{thebibliography}
\bibliographystyle{spiebib} 



\vspace{2ex}\noindent\textbf{Barry Lyu} is an Electrical and Computer Engineering PhD student at the University of Michigan. He received his BS degree in Computer Science from Cornell University in 2025. 

\vspace{2ex}\noindent\textbf{Vaibhavi Manjarekar} is an electrical engineering undergraduate at the University of Michigan. 

\vspace{2ex}\noindent\textbf{Nathaniel Bleier} is an assistant professor at the University of Michigan in the Computer Science and Engineering (CSE) department. He received his PhD in Electrical and Computer Engineering from the University of Illinois at Urbana-Champaign (UIUC) in 2024.

\vspace{1ex}
\noindent Biographies and photographs of the other authors are not available.

\listoffigures
\listoftables

\end{spacing}
\end{document}